\documentstyle[aps,pre,amssymb,amsmath,epsfig]{revtex}

\begin{document}


\title{Constructing, characterizing, and simulating  Gaussian and 
higher--order point distributions}
\author{Martin Kerscher\thanks{email: kerscher@theorie.physik.uni-muenchen.de}}
\address{
Sektion Physik, Ludwig--Maximilians--Universit{\"a}t, 
Theresienstra{\ss}e 37, D--80333 M{\"u}nchen, Germany\\
Department of Physics and Astronomy, 
The Johns Hopkins University, Baltimore, MD 21218, USA}
\date{submitted May 28, 2001; accepted July 31, 2001}
\maketitle

\begin{abstract}
The definition and the properties of a Gaussian point distribution, in
contrast to the well--known properties of a Gaussian random field are
discussed.  Constraints for the number density and the two--point
correlation function arise.  A simple method for the simulation of
this so--called Gauss--Poisson point process is given and illustrated
with an example.  The comparison of the distribution of galaxies in
the PSCz catalogue with the Gauss--Poisson process underlines the
importance of higher--order correlation functions for the description
for the galaxy distribution.  The construction of the Gauss--Poisson
point process is extended to the $n$--point Poisson cluster process,
now incorporating correlation functions up to the $n$th--order.  The
simulation methods and constraints on the correlation functions are
discussed for the $n$--point case and detailed for the three--point
case.  As another approach, well suited for strongly clustered
systems, the generalized halo--model is discussed. The influence of
substructure inside the halos on the two-- and three--point
correlation functions is calculated in this model.
\end{abstract}

%
%
%
%

\pacs{02.50.-r 02.30.Mv 02.70.Rr 98.65.-r}

\twocolumn
\narrowtext

\section{Introduction}

Stochastic models are often used to describe physical phenomena.  For
spatial structures two broad classes of stochastic models have been
established.  One approach is based on random fields and the other one
on random distributions of discrete objects, often only points, in
space (see the contributions in {}\cite{mecke:statistical} for
recent applications and reviews).  Stochastic point distributions are
used to describe physical systems on vastly differing length--scales.
The physical applications discussed in this article deal with the
large--scale structures in the Universe formed by the distribution of
galaxies. However, the methods are much more versatile.

Models for the dynamics of cosmic structures are often based on
nonlinear partial differential equations for the mass density and
velocity field.  These models relate the initial mass density and
velocity field, primarily modeled as Gaussian random fields, to the
present day values of these fields.  The nonlinear evolution leads to
non--Gaussian features in the fields.  However, observations supply us
with the positions of galaxies in space.  To compare theories with
observations one has to relate fields with point distributions. Both
deterministic or stochastic models have been used for this purpose so
far (e.g.\ {}\cite{bardeen:gauss,dekel:stochastic}).

Pursuing a direct approach, the observed spatial distribution of
galaxies (galaxy clusters etc.)  is compared to models for random
point sets.  Only a few attempts towards a dynamics of galaxies as
discrete objects have been made (see e.g.\ {}\cite{peebles:tracing}),
however stochastic models are quite common.  Following the works by
{}\cite{neyman:statistical,soneira:computer,white:hierarchy,balian:I},
and {}\cite{szapudi:higher} a purely stochastic description of the
spatial distribution of galaxies as points in space is given in this
article.

Models for stochastic point processes can be constructed using the
physical interactions of the objects, typically leading to Gibbs
processes (see e.g.\ {}\cite{widom:model,baddeley:area}, and the
generalizations by
{}\cite{likos:interfaces,baddeley:quermass,kendall:quermass}).
Another approach to construct point processes is based on purely
geometrical considerations, e.g.\ points are randomly distributed on
randomly placed line--segment (see
{}\cite{stoyan:stochgeom,buryak:correlation}).  As a third possibility
one can start from the characterization of point processes by the
probability generating functional (p.g.fl.) and the expansion in terms
of correlation functions.  This is the way pursued in this work.

The simplest point process is a Poisson process showing no
correlations at all. Since the galaxy distribution is highly
clustered, a Poisson process is not a realistic model. The model with
the next level of sophistication is a Gauss--Poisson process, the
point distribution counterpart of a Gaussian random field. Whereas the
properties of Gaussian random fields have been extensively studied,
the Gauss--Poisson process has not been discussed in the cosmological
literature in a systematic way.  Some of the statistical properties of
random fields directly translate to similar statistical properties of
point distributions, but also important differences show up.  The
systematic inclusion of higher--order correlations, as well as the
characterization, and the simulation algorithms for such point
processes will be discussed.

Recently, a related class of stochastic models for the galaxy
distribution, the halo--model, attracted some attention (see e.g.\
{}\cite{sheth:non-linear,ma:deriving,peacock:halo,scoccimarro:howmany}).
These models are based on the assumption that galaxies are distributed
inside correlated dark matter halos.  Using the probability generating
functional, the two-- and three--point correlation function will be
calculated for this model, extending the results by
{}\cite{scherrer:statistics} to include the effects of
halo--substructure.

The outline of this paper is as follows:\\ 
In Sect.~\ref{sect:pgfl} the properties of the probability generating
functional (p.g.fl.)  of a point process and the expansions in several
types of correlation functions are briefly reviewed.
The characterization of the Gauss--Poisson process is given in
Sect.~\ref{sect:gauss-poisson}, and the physical consequences of the
constraints are discussed. The close relation to Poisson cluster)
processes allows us to simulate a Gauss--Poisson process (see
Appendix~\ref{sect:simulate-Gauss}.
In Sect.~\ref{sect:examples} simulations of the Gauss--Poisson
processes and the line--segment process are used to show how the
Gaussian approximation influences the $J(r)$-function, a statistic
sensitive to higher--order correlations. A comparison of the galaxy
distribution within the PSCz survey with a Gauss--Poisson processes
illustrates the importance of higher--order correlations.
In Sect.~\ref{sect:higher-order} the extension of the Gauss--Poisson
point process to the $n$--point Poisson cluster process is discussed.
Detailed results are derived for the three--point Poisson cluster
process (the simulation recipe is give in
Appendix~\ref{sect:simul-threepoint}).  The characterization of the
general $n$--point process is discussed which is again detailed for
the three--point case.
Differences between a point process and a random field are
highlighted in Sect.~\ref{sect:field-point}.
Models for strongly correlated systems are mentioned in
Sect.~\ref{sect:strongly}. The focus will be on the ``halo model''.
Using the formalism based on the p.g.fl., the correlation functions of
the ``halo model'', including the effects of halo--substructure,
are calculated in Sect.~\ref{sect:beyond-halo}.
In Sect.~\ref{sect:problems} some open problems are mentioned.
An outlook is provided in Sect.~\ref{sect:summary}.
As an example the probability generating function (p.g.f.)  of a
random variable and its expansions in several kinds of moments is
reviewed in Appendix~\ref{sect:random-variables}.

\section{Product densities, factorial cumulants, and the 
probability generating functional}
\label{sect:pgfl}

Probability generating functionals (p.g.fl.'s), and their expansions
in different kinds of correlation measures have been used to describe
noise in time series (e.g.\ {}\cite{stratonovich:topicsI}) and the
electro--magnetic cascades occurring in air--showers (e.g.\
{}\cite{srinivasan:stochastic}).  They have been employed in the
theory of liquids (e.g.\ {}\cite{hansen:theory}) and other branches of
many--particle physics (e.g.\ {}\cite{ruelle:statistical}).  The
mathematical theory of p.g.fl.'s for point processes is nicely
reviewed in the book of {}\cite{daley:introduction}.  Stochastic
methods based on p.g.f.'s have been introduced to cosmology by
{}\cite{neyman:statistical} (the p.g.fl.\ was presented by
{}\cite{moyal:discussion} in the discussion of this article), and
became well--known following the work of {}\cite{white:hierarchy} and
{}\cite{balian:I}.  Focusing on the factorial moments (the volume
averaged $n$--point densities) and on count--in--cells,
{}\cite{szapudi:higher} discussed several expansions of the p.g.f.'s.
In the following only ``simple'' point processes will be considered:
at each position in space at most one object is allowed.  This
assumption is physically well justified for galaxies.  Also, for
quantum systems the methods should be refined (see e.g.\
{}\cite{soshnikov:determinantal} for fermionic (determinantal) point
processes).

An intuitive way to characterize a point process is to use
$n$th--product densities: $\varrho_n({\mathbf{x}}_1,\ldots,{\mathbf{x}}_n){\rm d}
V_1\cdots{\rm d} V_n$ is the probability of finding a point in each of
the volume elements ${\rm d} V_1$ to ${\rm d} V_n$.  For stationary and
isotropic point fields $\varrho_1({\mathbf{x}})=\varrho$ is the mean number
density, and the product density (with a slight abuse of notation) is
$\varrho_2({\mathbf{x}}_1,{\mathbf{x}}_2)=\varrho_2(r)$ with $r=|{\mathbf{x}}_1-{\mathbf{x}}_2|$ being the
separation of the two points.
The   factorial   cumulants   $c_{[n]}({\mathbf{x}}_1,\ldots,{\mathbf{x}}_n)$  are   the
irreducible  or  connected  parts  of  the  $n$th--product  densities.
E.g.\ for $n=2$ 
\begin{equation}
\varrho_2(r) = \varrho^2+c_{[2]}(r) = \varrho^2\Big(1+\xi_2(r)\Big),
\end{equation}
and the  second   factorial  cumulant  $c_{[2]}(r)$   and  the  two--point
correlation function  $\xi_2(r)$ quantify the  two--point correlations
in  excess  of  Poisson  distributed points.

A  systematic  characterization of a point process is provided by the
probability generating functional or a series of probability
generating functions (see Appendix~\ref{sect:random-variables}).
Equivalent to a random distribution of points in space, one considers
a point process as a random counting measure.  A realization is then a
counting measure $N$, which assigns to each suitable set
$A\subset{\mathbb{R}}^d$ the number of points $N(A)$ inside.
For suitable functions $h({\mathbf{x}})$ one defines the {\em probability
generating functional} of a point process via
\begin{equation}
G[h] = {\mathbb{E}}\Bigg[
\exp\Big(\int_{{\mathbb{R}}^d}N({\rm d} {\mathbf{x}}) \log h({\mathbf{x}}) \Big) \Bigg] ,
\end{equation}
where ${\mathbb{R}}^d$ is the $d$--dimensional Euclidean space, and ${\mathbb{E}}$ is
the expectation value, the ensemble average over realizations of the
point process.
Equivalently,
\begin{equation}
\label{eq:def-pgfl}
G[h] = {\mathbb{E}}\Big[\prod\nolimits_i h({\mathbf{x}}_i)\Big] ,
\end{equation}
where ${\mathbf{x}}_i$ are the particle positions in a realization.
Consider $k$  compact  disjoint  sets  $A_j$, and  let
$n_j=N(A_j)$ be the number of points inside $A_j$.  The p.g.f.\ of the
$k$-dimensional random vector $(n_1,\cdots,n_k)$ is then
\begin{equation}
\label{eq:gen-function}
P_k({\mathbf{z}})=P_k(z_1,\ldots,z_k)= {\mathbb{E}}\bigg[\prod_{j=1}^k z_j^{n_j}\bigg] .
\end{equation}
Together with a continuity requirement the knowledge of {\em all}
finite dimensional p.g.f.'s $P_k$ determines the p.g.fl.\ $G$ and the
point process completely (e.g.\ {}\cite{daley:introduction}).  One
obtains the p.g.f.\ $P_k({\mathbf{z}})=G[h']$ of the random vector ${\mathbf{z}}$ using
\begin{equation}
\label{eq:def-hstrich}
h'({\mathbf{x}}) = 1-\sum_{j=1}^k (1-z_j)\mbox{1\hspace*{-0.085cm}l}_{A_j}({\mathbf{x}}) ,
\end{equation}
where $\mbox{1\hspace*{-0.085cm}l}_A({\mathbf{x}})$ is the indicator--function of the set $A$, with
$\mbox{1\hspace*{-0.085cm}l}_A({\mathbf{x}})=1$ for ${\mathbf{x}}\in{A}$ and zero otherwise.
Several expansions of the p.g.fl.\ $G[h]$ are possible 
{}\cite{daley:introduction}.  The expansion in terms of the product 
densities $\varrho_n$ (the Lebesgue densities of the factorial moment 
measures) reads
\begin{multline}
\label{eq:fact-moment-expansion}
G[h+1] = 1 + \sum_{n=1}^\infty 
\frac{1}{n!}\int_{{\mathbb{R}}^d}{\rm d}{\mathbf{x}}_1\cdots\int_{{\mathbb{R}}^d}{\rm d}{\mathbf{x}}_n \\
\varrho_n({\mathbf{x}}_1,\ldots,{\mathbf{x}}_n)\ h({\mathbf{x}}_1) \cdots h({\mathbf{x}}_n) .
\end{multline}
For the factorial cumulants $c_{[n]}$ or correlation functions $\xi_n$
one obtains
\footnote{The relations to the generating functionals ${\cal{R}}$,
${\cal{F}}$ and ${\cal{G}}$ defined by {}\cite{balian:I} are
$G[h]={\cal{R}}[h]$, $G[h]={\cal{F}}[h-1]$ and
$G[h]=\exp{\cal{G}}[h-1]$.}
\begin{align}
\label{eq:cumulant-exp}
\log G[h+1]
&= \sum_{n=1}^\infty  
\frac{1}{n!}\int_{{\mathbb{R}}^d}{\rm d}{\mathbf{x}}_1\cdots\int_{{\mathbb{R}}^d}{\rm d}{\mathbf{x}}_n \nonumber\\
& \qquad c_{[n]}({\mathbf{x}}_1,\ldots,{\mathbf{x}}_n)\ h({\mathbf{x}}_1)\cdots h({\mathbf{x}}_n)\nonumber\\
&= \sum_{n=1}^\infty    
\frac{\varrho^n}{n!}\int_{{\mathbb{R}}^d}{\rm d}{\mathbf{x}}_1\cdots\int_{{\mathbb{R}}^d}{\rm d}{\mathbf{x}}_n \\
& \qquad \xi_n({\mathbf{x}}_1,\ldots,{\mathbf{x}}_n)\ h({\mathbf{x}}_1)\cdots h({\mathbf{x}}_n).\nonumber
\end{align}
As a third possibility the p.g.fl.\ can be expanded around the origin:
\begin{multline}
\label{eq:janossy-exp}
G[h] = J_0 + \sum_{n=1}^\infty   
\frac{1}{n!}\int_{{\mathbb{R}}^d}{\rm d}{\mathbf{x}}_1\cdots\int_{{\mathbb{R}}^d}{\rm d}{\mathbf{x}}_n \\
 j_n({\mathbf{x}}_1,\ldots,{\mathbf{x}}_n)\ h({\mathbf{x}}_1) \cdots h({\mathbf{x}}_n).
\end{multline}
The Janossy densities $j_n({\mathbf{x}}_1,\ldots,{\mathbf{x}}_n){\rm d} 
V_1\cdots{\rm d} V_n$ are the probability that there are exactly $n$ 
points, each in one of the volume elements ${\rm d} V_1$ to ${\rm d} V_n$.
Convergence issues of these expansions are discussed  in
Sect.~\ref{sect:problems}.

\section{The Gauss--Poisson point  process}
\label{sect:gauss-poisson}

For a stationary Poisson process with mean number density $\varrho$
the p.g.fl.\ is
\begin{equation}
\log G[h+1] = \varrho \int_{{\mathbb{R}}^d}{\rm d}{\mathbf{x}}\ h({\mathbf{x}}) ,
\end{equation}
corresponding to a truncation of the expansion~\eqref{eq:cumulant-exp}
after the first term.
Truncating after the second term, one obtains the p.g.fl.\ for the
Gauss--Poisson process {}\cite{newman:new,milne:further}
\begin{multline}
\label{eq:G-gauss}
\log G[h+1] = \varrho\int_{{\mathbb{R}}^d}{\rm d}{\mathbf{x}}\ h({\mathbf{x}}) +\\
+\frac{\varrho^2}{2}\int_{{\mathbb{R}}^d}{\rm d}{\mathbf{x}}\int_{{\mathbb{R}}^d}{\rm d}{\mathbf{y}}\ 
\xi_2(|{\mathbf{x}}-{\mathbf{y}}|)\ h({\mathbf{x}})h({\mathbf{y}}),
\end{multline}
completely  specified by  its mean  number density  $\varrho$  and the
two-point correlation function $\xi_2(r)$.

There is a close resemblance to random fields.  For a homogeneous
random field $\rho({\mathbf{x}})$ with mean ${\mathbb{E}}[\rho]=\overline{\rho}$ the
density contrast is defined by
$\delta({\mathbf{x}})=(\rho({\mathbf{x}})-\overline{\rho})/\overline{\rho}$.  A
homogeneous and isotropic Gaussian random field is stochastically
fully specified by its mean $\overline{\rho}$ and its correlation
function $\xi_2^\delta(|{\mathbf{x}}-{\mathbf{y}}|)={\mathbb{E}}[\delta({\mathbf{x}})\delta({\mathbf{y}})]$
{}\cite{adler:randomfields}.  Here, ${\mathbb{E}}$ is the average over
realizations of the random field.  The higher (connected) correlation
functions $\xi_n^\delta=0$ with $n>2$ vanish.  Similar the correlation
functions $\xi_n$ for $n>2$ vanish in a Gauss--Poisson process.
However,  also important  differences between  a Gaussian  {\em random
field} and a Gauss--Poisson {\em point process} show up.

\subsection{Constraints on $\xi_2(r)$ and $\varrho$}
\label{sect:constraints-gauss}

A functional $G[h]$ defined by Eq.~\eqref{eq:G-gauss} is a p.g.fl.\ of
a point process if and only if the $P_k({\mathbf{z}})$ as given in
Eq.~\eqref{eq:gen-function} are probability generating functions
(p.g.f.'s).  This will lead to restrictions on the two--point
correlation function $\xi_2$ and the number density $\varrho$ as
discussed by {}\cite{newman:new} and {}\cite{milne:further}.  A
$P_k({\mathbf{z}})$ given by Eq.~\eqref{eq:gen-function} always has to be
positive and monotonic increasing with each component $z_i$ of ${\mathbf{z}}$,
and hence $\log P_k({\mathbf{z}})$ is non--decreasing in each component of
${\mathbf{z}}$.  With Eqs.~\eqref{eq:gen-function}, {}\eqref{eq:def-hstrich}
and {}\eqref{eq:G-gauss} one gets
\begin{multline}
\label{eq:const-general}
\frac{\partial\log P_k({\mathbf{z}})}{\partial z_l} =
\varrho|A_l| + \\
+\varrho^2\sum_{j=1}^k \int_{A_l}{\rm d}{\mathbf{x}}\int_{A_j}{\rm d}{\mathbf{y}}\
\xi_2(|{\mathbf{x}}-{\mathbf{y}}|)(z_j-1) \ge 0
\end{multline}
for any $z_j\ge0$, where $|A_l|$ is the volume of the set $A_l$.
The rather obvious constraint $\varrho\ge0$ can be derived by setting
$z_j=1$.  For $z_j=1,j\ne i$, and either $z_i=0$ or $z_i\gg1$ the
following two non--trivial constraints emerge:
\begin{gather}
\label{eq:const1}
\frac{\varrho}{|A_l|}\ \int_{A_l}{\rm d}{\mathbf{x}}\int_{A_i}{\rm d}{\mathbf{y}}\ \xi_2(|{\mathbf{x}}-{\mathbf{y}}|) 
\le 1 \\
\label{eq:const2}
\int_{A_l}{\rm d}{\mathbf{x}}\int_{A_i}{\rm d}{\mathbf{y}}\ \xi_2(|{\mathbf{x}}-{\mathbf{y}}|) 
\ge \frac{-|A_l|}{\varrho(z_i-1)} \overset{z_i\rightarrow\infty}{\longrightarrow} 0 .
\end{gather}
One can show that these two conditions provide a necessary and
sufficient characterization of $\xi_2(r)$ and $\varrho$, to assure
that $G[h]$, as given in Eq.~\eqref{eq:G-gauss}, is a p.g.fl.\
{}\cite{milne:further}.
 
Eq.~\eqref{eq:const1} constrains the shape and normalization of the
two--point correlation functions admissible in a Gauss--Poisson
process.  For $A_l=A=A_i$
\begin{equation}
\label{eq:const1a}
\frac{\varrho}{|A|}\ \int_{A}{\rm d}{\mathbf{x}}\int_{A}{\rm d}{\mathbf{y}}\ \xi_2(|{\mathbf{x}}-{\mathbf{y}}|)
=  \overline{N}\ \sigma^2(A) \le 1,
\end{equation}
where $\sigma^2(A)$ are the fluctuations  of count--in--cells in excess
of  a Poisson process,  and $\overline{N}=\varrho  |A|$ is  the mean
number of points  inside the cell $A$.  Hence,  the total fluctuations
of the  number of points $N$  inside $A$ for  a Gauss--Poisson process
are
\begin{equation}
{\mathbb{E}}[(N-\overline{N})^2] = \overline{N} + \overline{N}^2\ \sigma^2(A) 
\le 2\ \overline{N},
\end{equation}
and must not exceed twice the value of the fluctuations in a Poisson
process ($\sigma^2=0$) for any domain $A$ considered.
Another way of looking at constraint~\eqref{eq:const1} is by taking
$A_l$ as an infinitesimal volume element centered on the origin and
$A_j$ equal to some volume $A$:
\begin{equation}
\label{eq:const2-alternate}
\varrho\ \int_{A}{\rm d}{\mathbf{y}}\ \xi_2(|{\mathbf{y}}|) \le 1.
\end{equation}
Consistent with Sect.~\ref{sect:gp-as-pc} this tells us that sitting 
on a point of the process on average at most one other point in excess 
of Poisson distributed points can be present.

Now consider two volume elements $A_i={\rm d} V_i$ and $A_l={\rm d} V_l$
separated by a distance of $r$, then Eq.~\eqref{eq:const2} implies
\begin{equation}
\xi_2(r)\ge0
\end{equation}
Hence, only clustered point distributions can be modeled by a
Gauss--Poisson process.  Any zero crossing in $\xi_2(r)$ already
indicates the presence of higher--order correlations.  

\subsection{A Gauss--Poisson process as a Poisson cluster process}
\label{sect:gp-as-pc}

A Gauss--Poisson process can be interpreted as a simple Poisson
cluster process.  This is important for simulations (see
Appendix~\ref{sect:simulate-Gauss}).

A Poisson cluster process is a two--stage point process.  First one
chooses Poisson distributed cluster centers, the ``parents'', with
number density $\varrho_p$ and then attaches a second point process --
the cluster to each cluster center (the cluster center is not
necessarily part of the point process).  The p.g.fl.\ of a Poisson
cluster process is then given by {}\cite{daley:introduction}
\begin{equation}
\label{eq:poisson-cluster-G}
\log G[h] = \int_{{\mathbb{R}}^d}{\rm d}{\mathbf{x}}\ \varrho_p \big(G_c[h|{\mathbf{x}}]-1\big) ,
\end{equation}
with $G_c[h|{\mathbf{x}}]$ being the p.g.fl.\  of the point process forming the
cluster at center ${\mathbf{x}}$.
Now  consider the p.g.fl.\  of a  cluster with  at maximum  two points
(compare with Eq.~\eqref{eq:def-pgfl}),
\begin{equation}
\label{eq:Gauss-poisson-cluster}
G_c[h|{\mathbf{x}}] = q_1({\mathbf{x}})h({\mathbf{x}}) + 
q_2({\mathbf{x}})h({\mathbf{x}})\int_{{\mathbb{R}}^d}{\rm d}{\mathbf{y}} f(|{\mathbf{x}}-{\mathbf{y}}|)\ h({\mathbf{y}}) ,
\end{equation}
where $q_1({\mathbf{x}})$ is the probability that only one point, the cluster
center at ${\mathbf{x}}$, is entering the cluster, whereas $q_2({\mathbf{x}})$ is the
probability that a second point is added.  Clearly,
$q_1({\mathbf{x}})+q_2({\mathbf{x}})=1$.  The probability density $f(|{\mathbf{x}}-{\mathbf{y}}|)$
determines the distribution of the distance $|{\mathbf{x}}-{\mathbf{y}}|$ of the second
point ${\mathbf{y}}$ to the cluster center, and is normalized according to
$\int{\rm d}{\mathbf{z}}{}f(|{\mathbf{z}}|)=1$.
By writing $f(|{\mathbf{x}}-{\mathbf{y}}|)$ one assumes that the probability density $f$
is symmetric in ${\mathbf{x}}$ and ${\mathbf{y}}$.  Indeed, the p.g.fl.\
Eq.~\eqref{eq:poisson-cluster-G} is invariant under interchanging
${\mathbf{x}}$ and ${\mathbf{y}}$, and this assumption does not impose any restrictions.
Using this expression and Eq.~\eqref{eq:poisson-cluster-G} one obtains
\begin{multline}
\log G[h+1] = \int_{{\mathbb{R}}^d}{\rm d}{\mathbf{x}}\ \varrho_p(1+q_2({\mathbf{x}}))\ h({\mathbf{x}}) +\\
+\int_{{\mathbb{R}}^d}{\rm d}{\mathbf{x}}\int_{{\mathbb{R}}^d}{\rm d}{\mathbf{y}}\ 
 \varrho_p q_2({\mathbf{x}}) f(|{\mathbf{x}}-{\mathbf{y}}|)\ h({\mathbf{x}})h({\mathbf{y}}),
\end{multline}
which    equals    the     p.g.fl.\    for    the    Gauss--Poisson
process~\eqref{eq:G-gauss}    for    $\varrho=\varrho_p(1+q_2)$    and
$\xi_2(r)=2\frac{\varrho-\varrho_p}{\varrho^2}f(r)$.     Hence    {\em
every} Gauss--Poisson  process is a Poisson cluster  process of the
above type, and vice versa.

\subsection{Physical implications}
\label{sect:physical}

From the preceding section one concludes that at maximum two points
form a cluster in a Gauss--Poisson process. Therefore, no point
distribution with large--scale structures can be modeled reliably with
this kind of process.  This has physical implications both for the
galaxy distribution and percolating/critical systems.

More specific, from the observed galaxy distribution a
scale--invariant two--point correlation function
$\xi_2(r)=Ar^{-\gamma}$ with $\gamma\approx1.8$ is deduced.  Clearly
such a correlation function does not satisfy the
constraint~\eqref{eq:const1}.  For $0<\gamma\le3$ a cut--off at large
scales has to be introduced.
For the galaxy distribution a cut--off at approximately 20$h^{-1}$Mpc is
the lowest value which is roughly compatible with the observed
two--point correlation function.  Taking into account the observed
number density of the galaxies, a cut--off even on this small scale
does not help. Still the constraint~\eqref{eq:const1} is strongly
violated, indicating non negligible higher--order correlation
functions (see also Sect.~\ref{sect:observed-galaxy}).
Similarly, a zero crossing or a negative $\xi_2(r)$ is violating the
constraint~\eqref{eq:const2} and also indicates that higher--order
correlations are present.  There are indications that the distribution
of galaxies shows a negative $\xi_2(r)$ on some scale larger than
20$h^{-1}$Mpc, followed by a positive peak at approximately
120$h^{-1}$Mpc
{}\cite{broadhurst:large-scale,mo:typical_scales,einasto:120mpc}. A
Gauss--Poisson process is not able to describe these features in the
distribution of galaxies and galaxy clusters.

Also a percolating cluster shows scale--invariant correlations.  The
correlation length, specifying e.g.\ the exponential cut--off of the
two--point correlation function, is going to infinity near the
percolation threshold.  Therefore, the geometry of the largest cluster
cannot be modeled with a Gauss--Poisson processes.  Higher--order
correlations are essential to describe the morphology of such a
system.  This again illustrates that the tails of the distributions,
in this case the asymptotic behavior of the two--point correlation
function is essential.

To summarize these results: already by looking at the two--point
correlation function and the density one is able to exclude a
Gauss--Poisson process as a model. However one cannot turn the
argument around and show that a given point distribution is compatible
with a Gauss--Poisson process using the two--point correlation
function alone. There are point processes with higher--order
correlations satisfying the constraints
({}\ref{eq:const1},\ref{eq:const2}) as discussed in
Sect.~\ref{sect:twod-example}.

\section{Detecting deviations from a Gauss--Poisson process}
\label{sect:examples}

After having outlined the basic theory of a Gauss--Poisson process, we
discuss in this section how one can detect non--Gaussian features in a
given point set.

\subsection{The line--segment process}
\label{sect:twod-example}

First a two dimensional analytic example is studied. In the
line--segment process points are randomly distributed on line segments
which are themselves uniformly distributed in space and direction.
The number of points per line segment is a Poisson random variable.
According to {}\cite{stoyan:stochgeom}, p.~286
\begin{equation}
\label{eq:xi2-line-seg}
\xi_2(r) = 
\begin{cases}
\frac{1}{\pi l \varrho_s} \big(\frac{1}{r}-\frac{1}{l}\big)
 & \text{ for } r < l\\
0 & \text{ for } r \ge l ;
\end{cases}
\end{equation}
$l$ is the length of the line segments and $\varrho_s$ is the mean
number density of line segments; $l\varrho_s$, $\varrho/\varrho_s$,
$\varrho$ denote the mean length density, the mean number of points
per line segment (which can be smaller than one), and the mean number
density in space, respectively.  A similar model for the distribution
of galaxies was discussed by {}\cite{buryak:correlation}.  On small
scales $r\ll l$, $\xi_2(r)\propto r^{-1}$, qualitatively similar to
the observed two--point correlation function in the galaxy
distribution.

\begin{figure}
\begin{center}
\epsfig{file=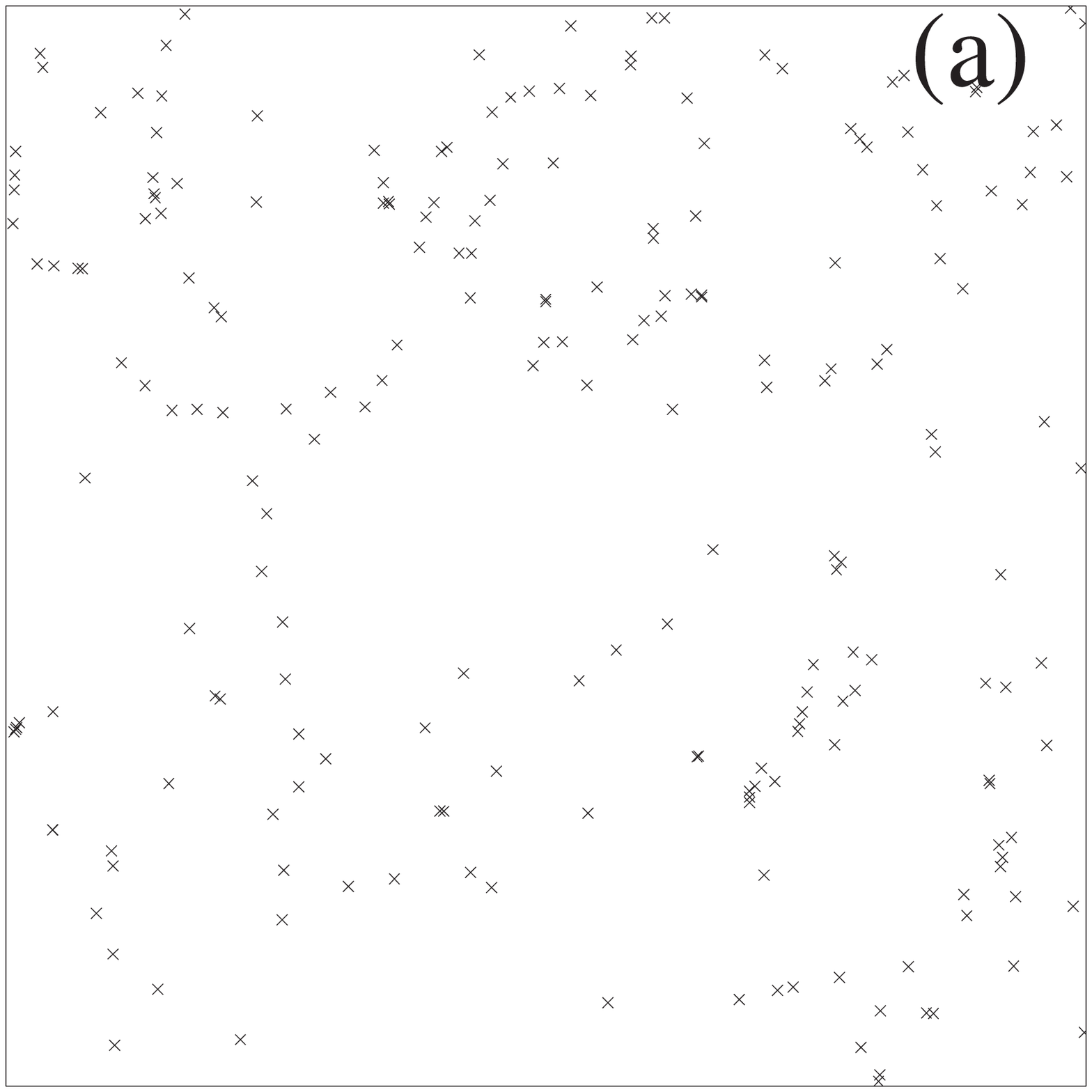,width=4.2cm}
\epsfig{file=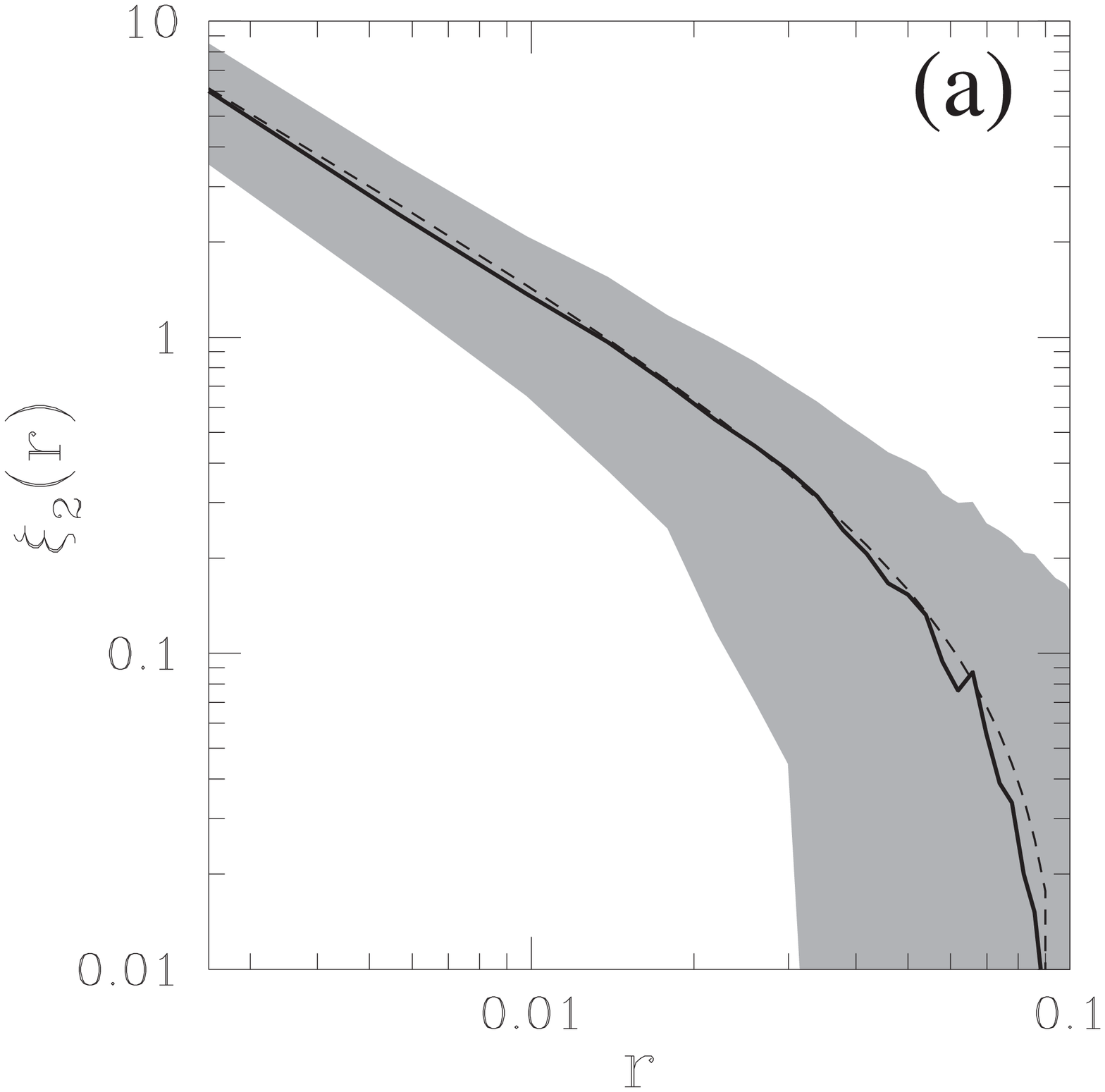,width=4.2cm}
\epsfig{file=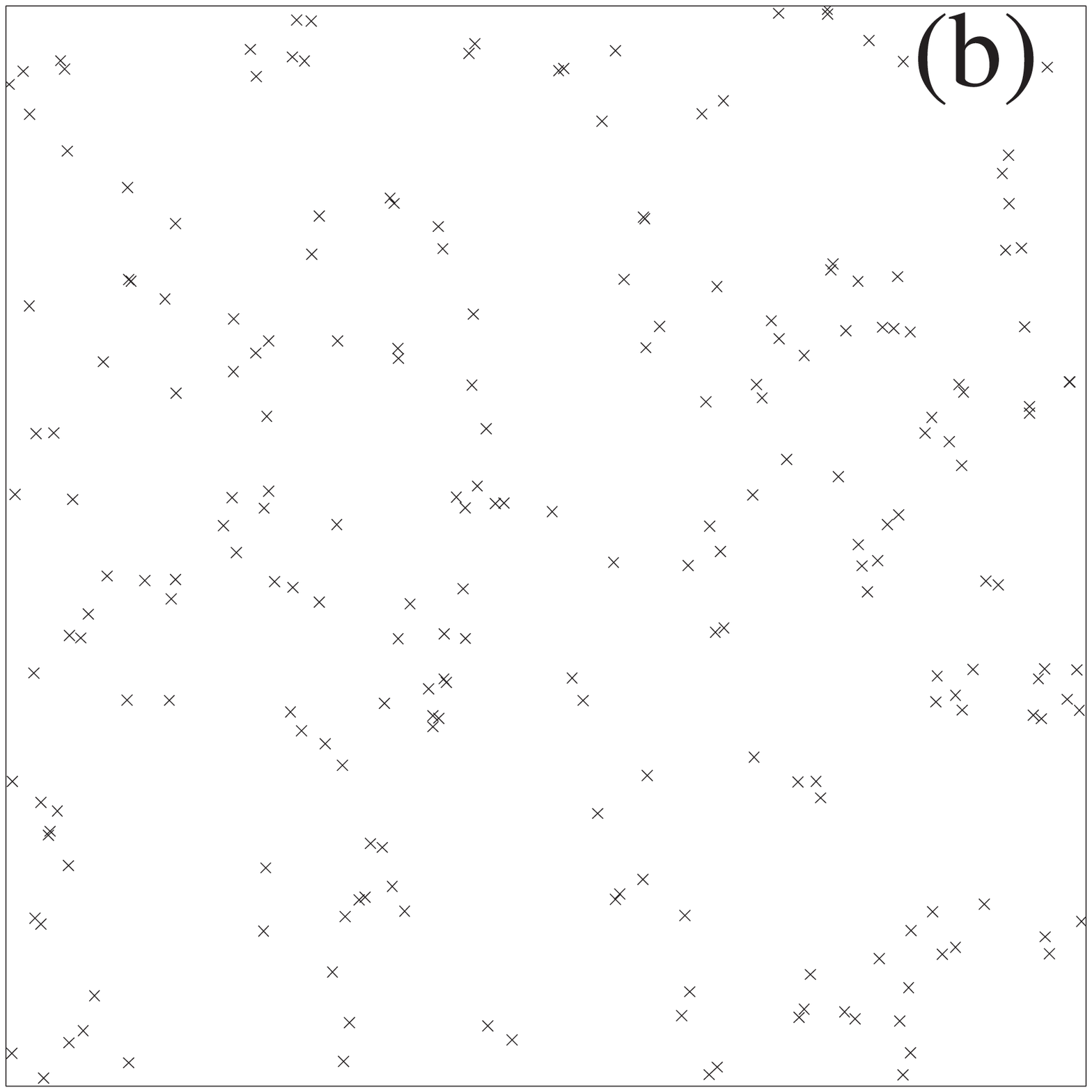,width=4.2cm}
\epsfig{file=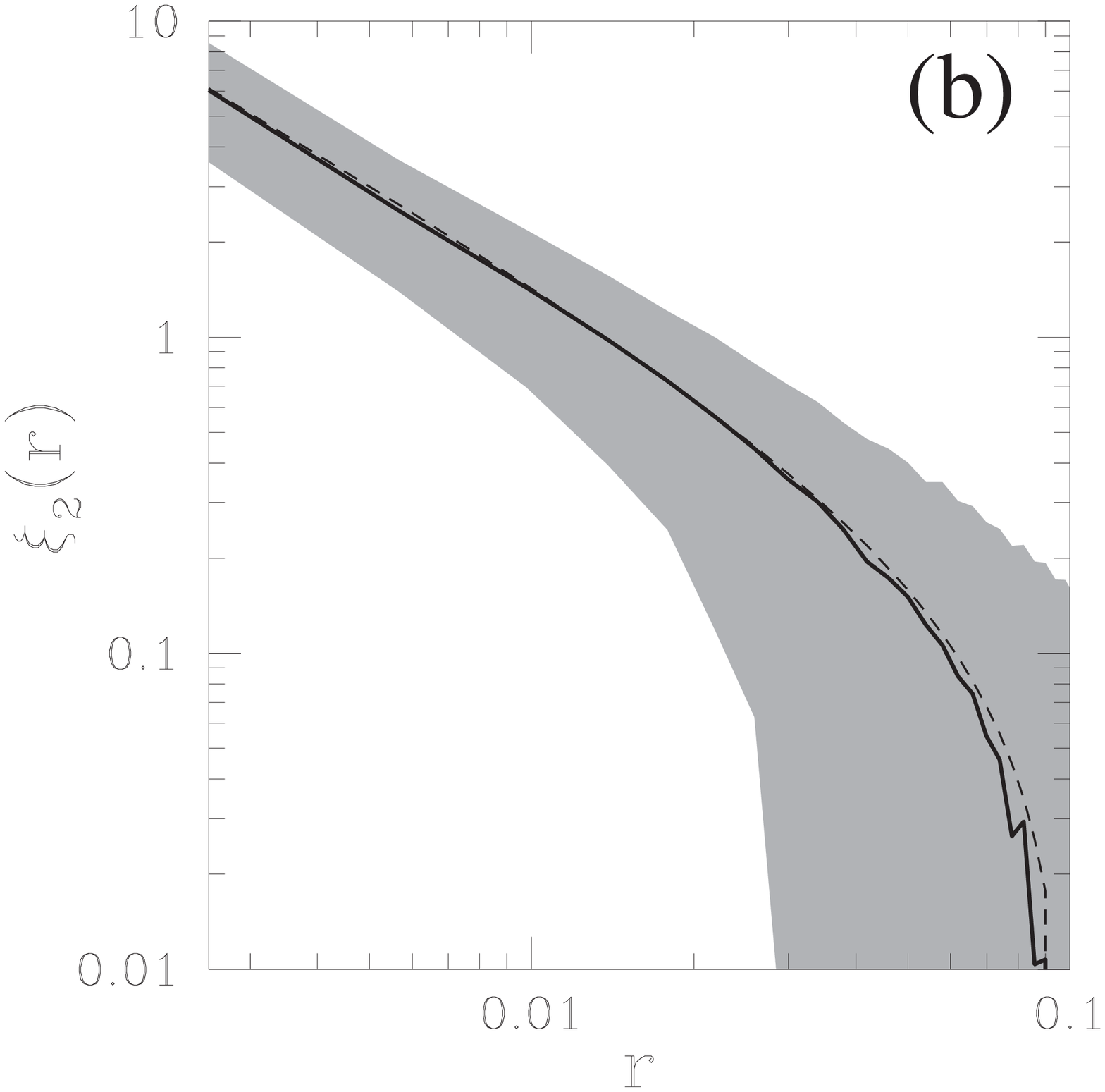,width=4.2cm}
\epsfig{file=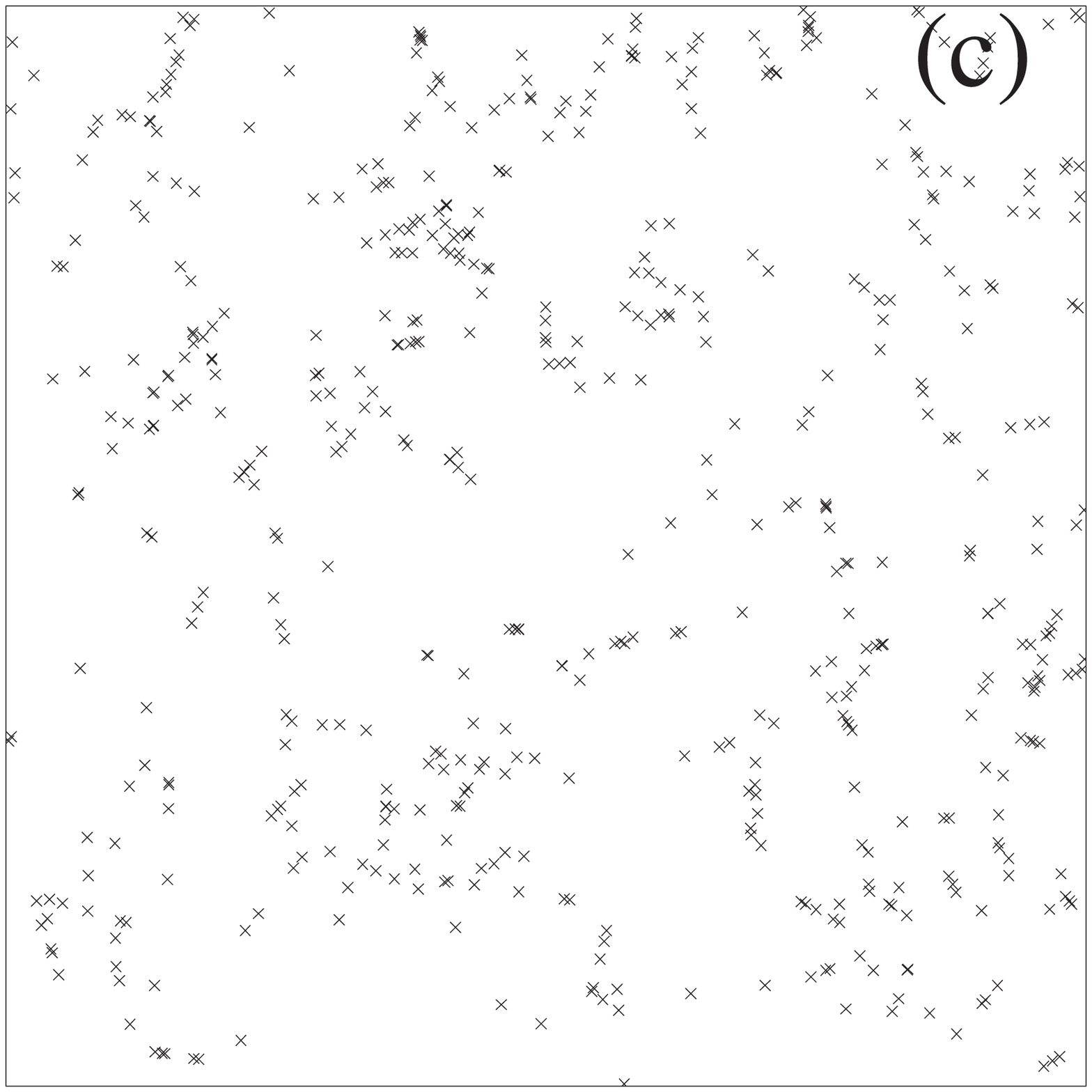,width=4.2cm}
\epsfig{file=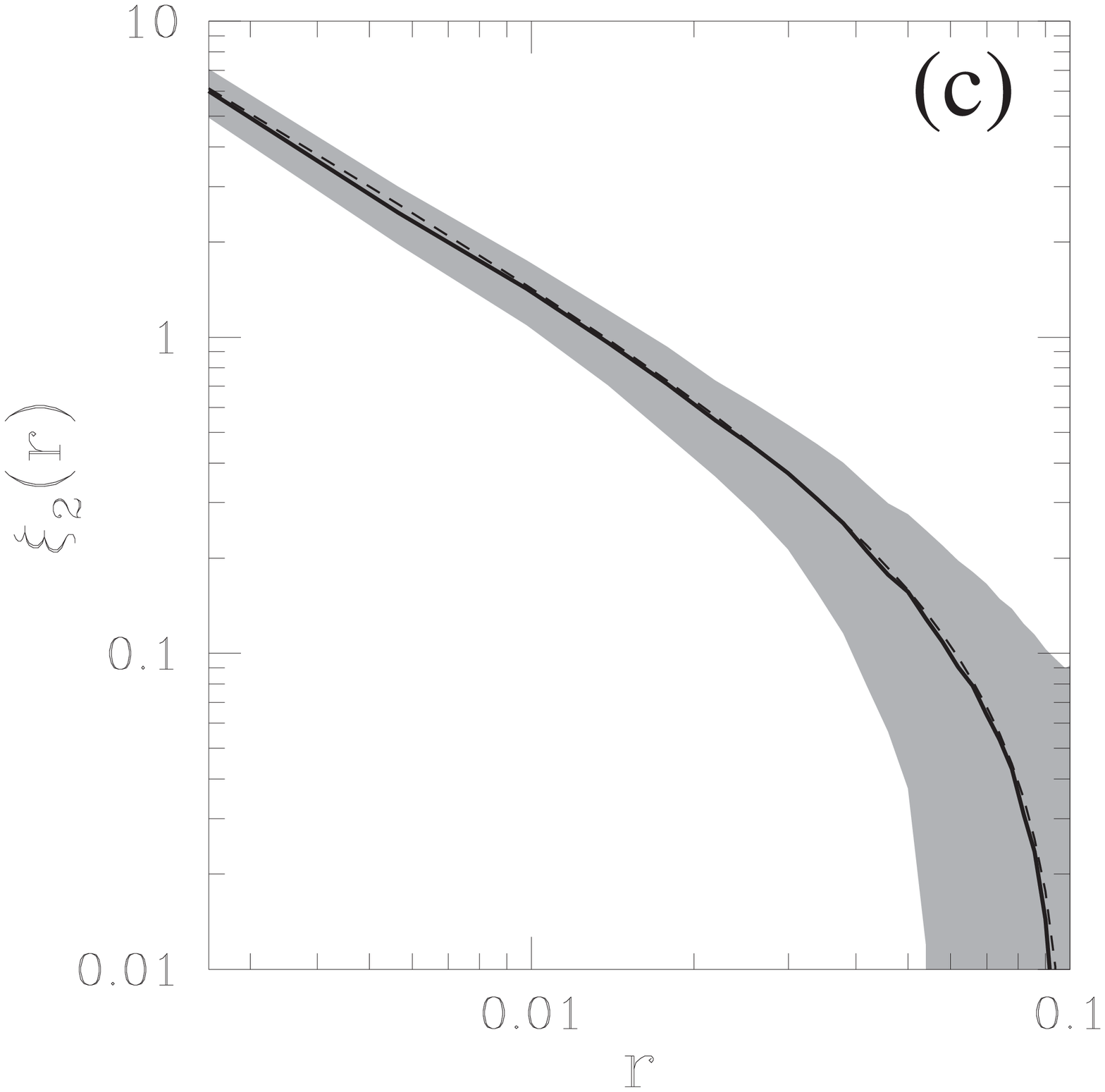,width=4.2cm}
\end{center}
\caption{\label{fig:line-gausspoi} The plot~(a) shows a realization
of the line--segment process inside the unit square with number
density $\varrho=200$ (for the other parameters see the text) and the
corresponding two--point correlation function $\xi_2(r)$ ($r$ is in
units of the box length): the dashed one--$\sigma$ area was determined
from 1000 realizations, the theoretical value is given by the dashed
line, nearly on top of the sample mean (solid line).  
The plot~(b) shows a realization of the Gauss--Poisson process with
$\varrho=200$ and the plot~(c) a realization of the high density
line--segment process with $\varrho=500$, both with the corresponding
two--point correlation function.}
\end{figure}

This structured point process incorporates higher--order correlations.  
In Fig.~\ref{fig:line-gausspoi} the line--segment process is shown in 
comparison to a Gauss--Poisson process with the same two--point 
correlation function for the parameters $l=0.1$, $\varrho_s=201$, 
$\varrho=200$, and $\varrho=500$.
A number density $\varrho>\varrho_s$ violates the constraint
Eq.~\eqref{eq:const1} and no Gauss--Poisson process equivalent on the
two--point level to such a line--segment process exists.

\subsection{Detecting higher--order correlations}
\label{sect:detect-highorder}

As can be seen from Fig.~\ref{fig:line-gausspoi}, the point processes
are indistinguishable on the two--point level. For another example see
{}\cite{baddeley:cautionary,arns:predicting}.  The differences between
these point distributions can be investigated with statistical methods
sensitive to higher--order correlations. One may use Minkowski
functionals ({}\cite{mecke:robust}, for reviews see
{}\cite{mecke:additivity,kerscher:statistical}), percolation
techniques {}\cite{shandarin:percolation}, the minimum spanning tree
{}\cite{barrow:minimalspanning}, a method sensitive to three--point
correlations {}\cite{schladitz:third}, or directly calculate the
higher moments
{}\cite{groth:statistical,fry:statistical,szapudi:cluster,szapudi:higher}.
In the following the $J$--function is used to quantify the
higher--order clustering
{}\cite{vanlieshout:j,kerscher:regular,kerscher:global}.

To define $J$-function the spherical contact distribution $F(r)$ is
needed, i.e.\ the {\em distribution function of the distances $r$
between an arbitrary point and the nearest object in the point set}.
$F(r)$ is equal to one minus the void--probability function:
$F(r)=1-P_0(r)$.  Another ingredient is the nearest neighbor distance
distribution $G(r)$, defined as the {\em distribution function of
distances $r$ of an object in the point set to the nearest other
point} {}\cite{hertz:ueber}.  For a Poisson process the probability to
find a point only depends on the mean number density $\varrho$,
leading to the well--known result
\begin{equation} \label{eq:FG_poi}
G_{\rm P}(r) = 1 - \exp\big(- \varrho |B_r|\big) 
= F_{\rm P}(r),
\end{equation}
where $|B_r|$ is  the volume of a $d$--dimensional  sphere with radius
$r$.
The ratio
\begin{equation}
J(r) = \frac{1-G(r)}{1-F(r)}
\end{equation}
was suggested by {}\cite{vanlieshout:j} as a probe for clustering of
a point distribution.  For a Poisson distribution $J(r)=1$ follows
directly from Eq.~\eqref{eq:FG_poi}.  A clustered point distribution
implies $J(r)\le1$, whereas regular structures are indicated by
$J(r)\ge1$.
As discussed in {}\cite{kerscher:regular} one can express the $J(r)$ 
function in terms of the $n$--point correlation functions $\xi_n$:
\begin{equation} 
J(r) = 1 + \sum_{l=1}^\infty \frac{(-\varrho)^l}{l!}
\int_{B_r}{\rm d}{\mathbf{x}}_1\dots\int_{B_r}{\rm d}{\mathbf{x}}_l\ 
\xi_{l+1}(0,{\mathbf{x}}_1,\ldots,{\mathbf{x}}_l) .
\end{equation}
$B_r$ is a $d$--dimensional sphere with radius $r$ centered on the
origin.
For a Gauss--Poisson process in two dimensions, i.e.\ $\xi_n=0$ for
$n>2$, the above expression simplifies\footnote{Unfortunately
{}\cite{kerscher:regular} discussed this Gaussian approximation with
examples of two--point correlation functions, which are not admissible
in a Gauss--Poisson process.}:
\begin{equation} \label{eq:Jgauss}
J(r) = 1 - \varrho\ 2\pi\ \int_0^r {\rm d} s\ s\ \xi_2(s).
\end{equation}

\begin{figure}
\begin{center}
\epsfig{file=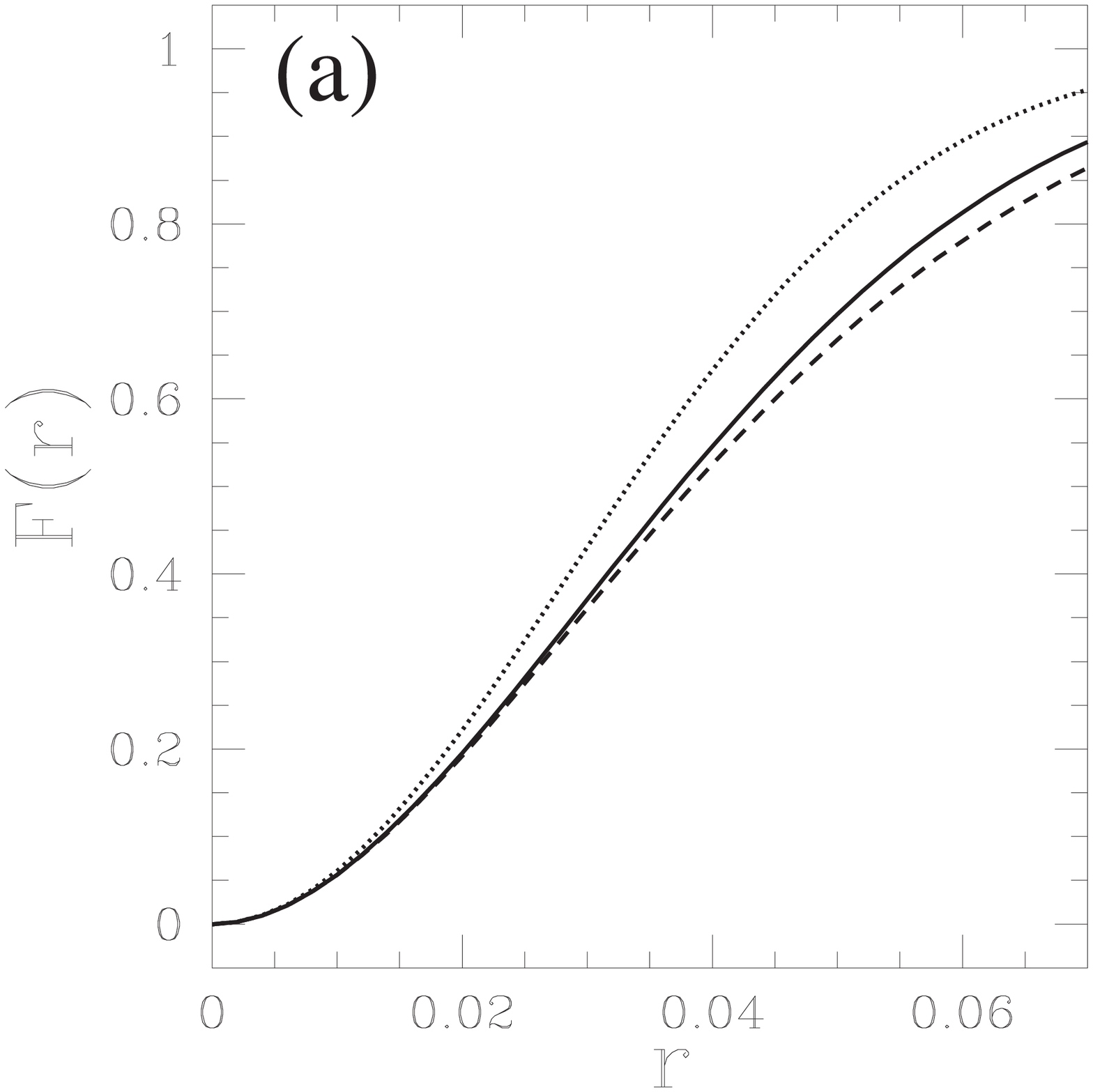,width=4.2cm}
\epsfig{file=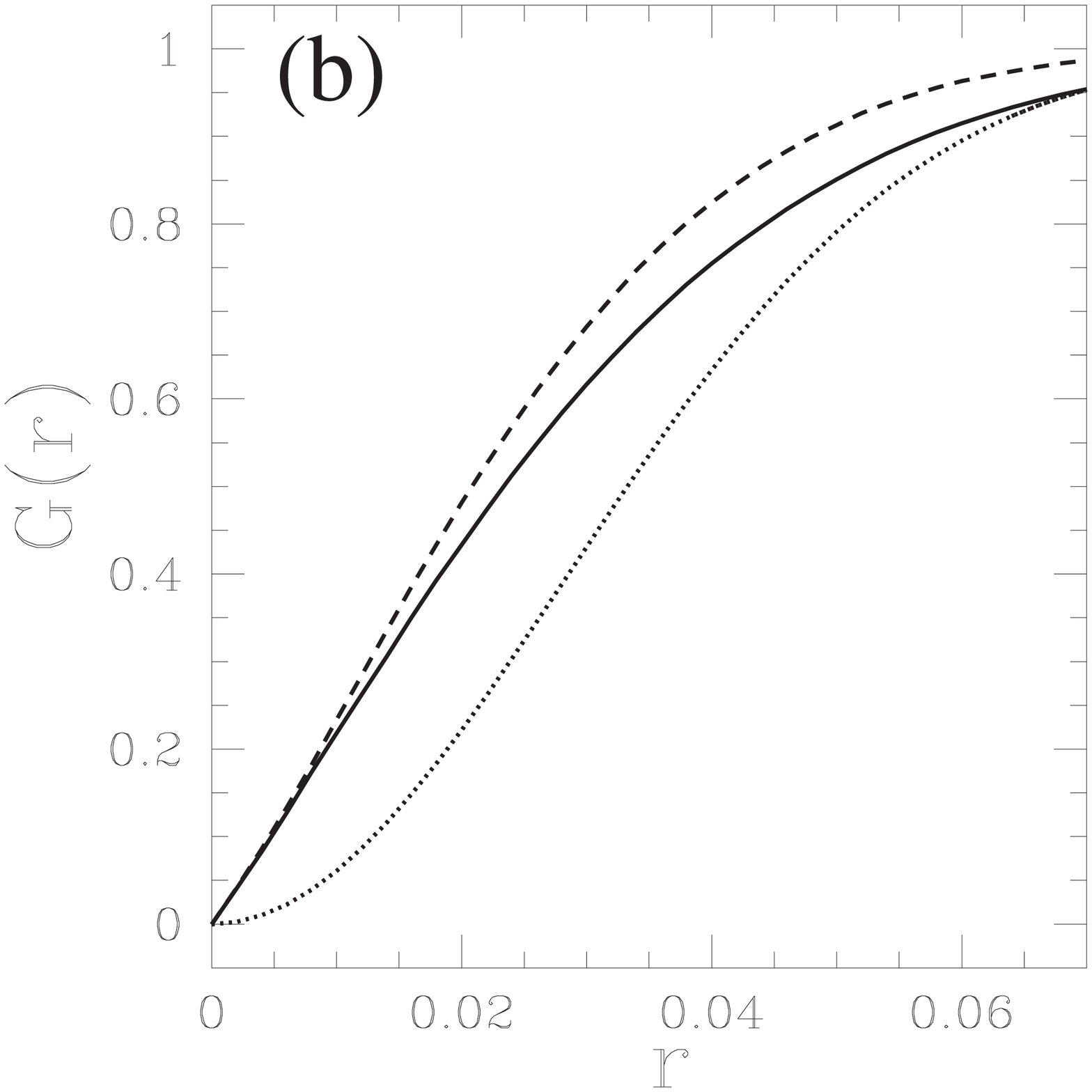,width=4.2cm}
\epsfig{file=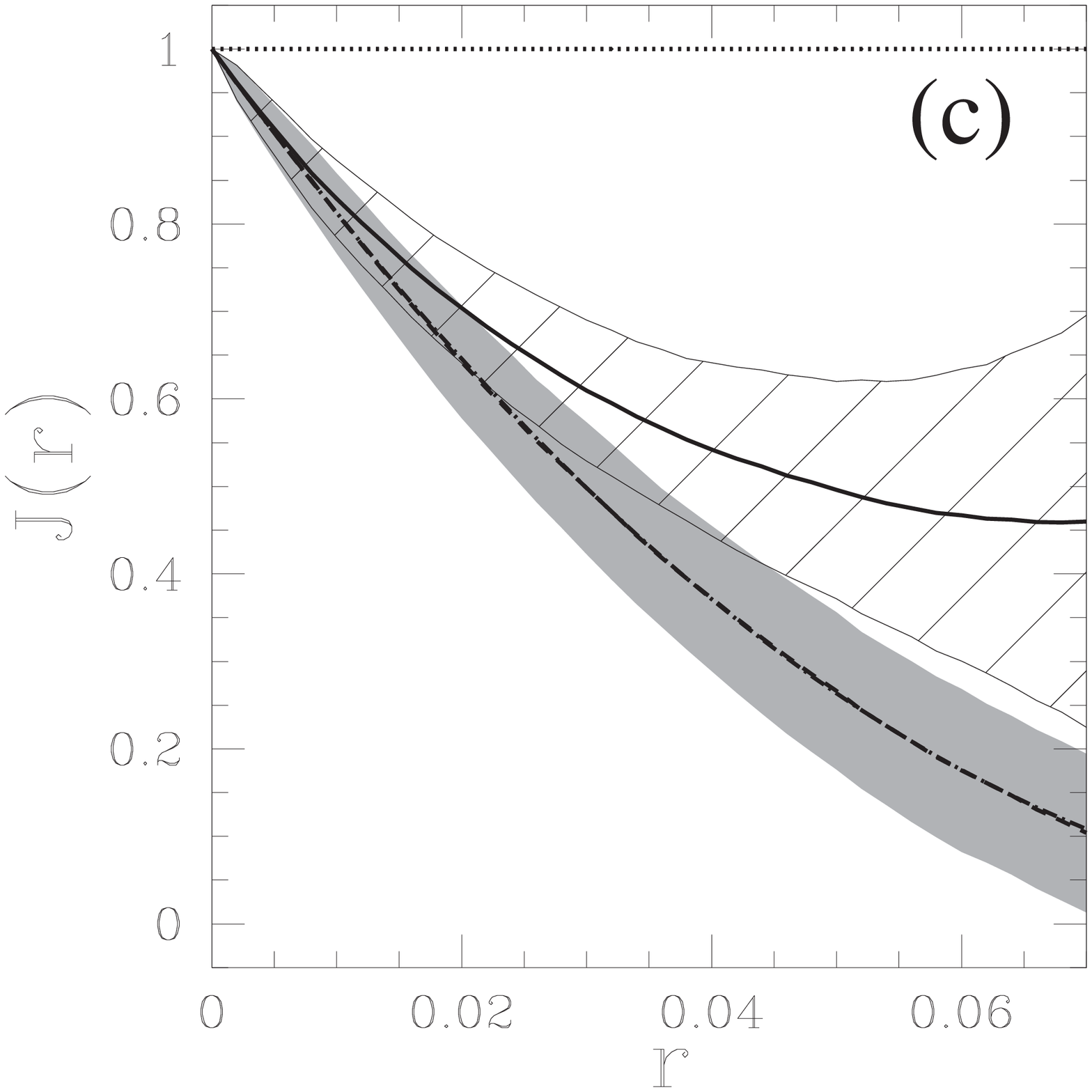,width=4.2cm}
\epsfig{file=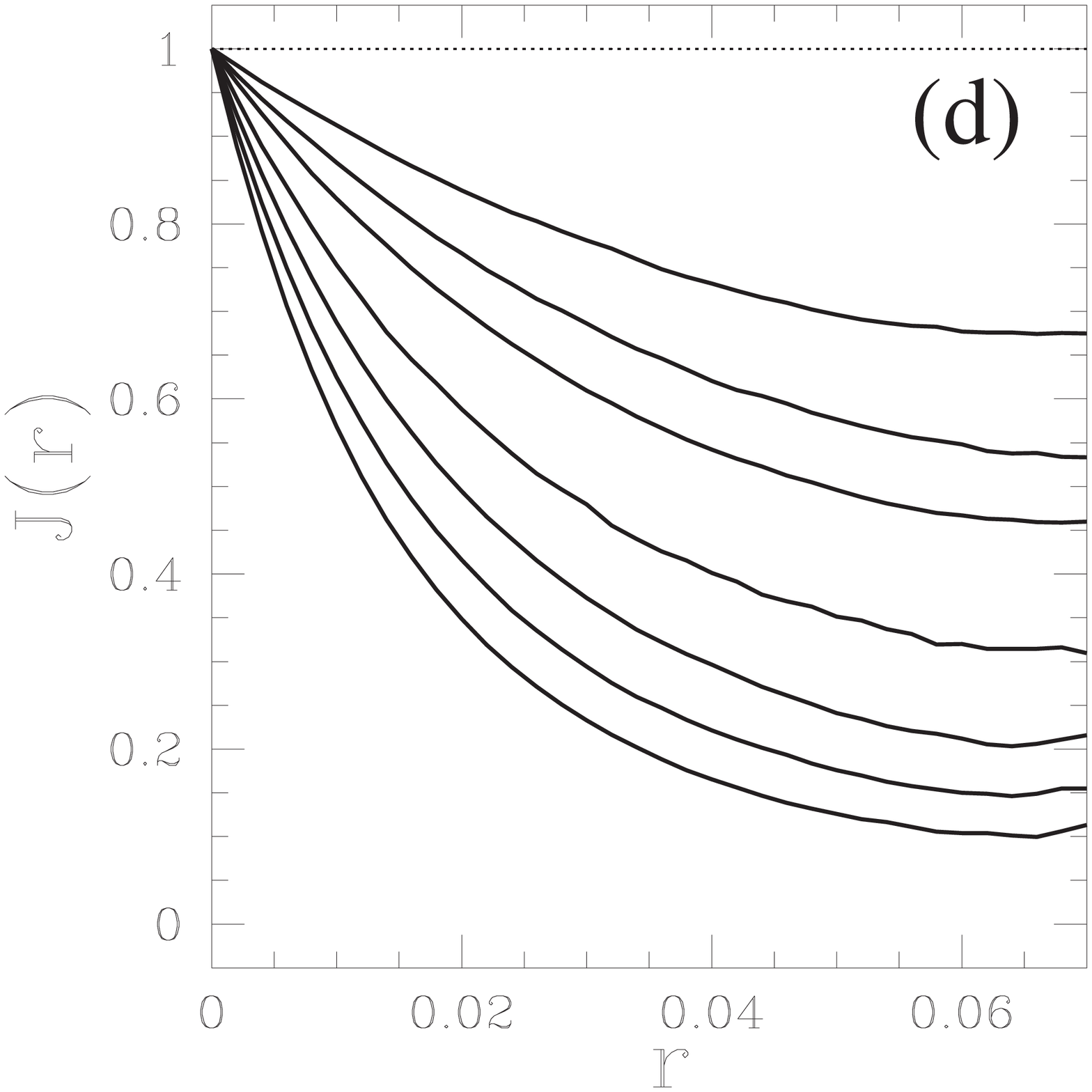,width=4.2cm}
\end{center}
\caption{\label{fig:FGJ-2d} The spherical contact distribution $F(r)$
(plot~(a)), the nearest neighbor distribution $G(r)$ (plot~(b)) and
the $J(r)$ function (plot~(c)) are shown for the Gauss--Poisson
process ($\varrho=200$, mean value (dashed line) and variance (shaded
area) estimated from 1000 realizations), and the line--segment process
($\varrho=200$, solid line); $r$ is in units of the box length. The
$J(r)$ function according to Eq.~\eqref{eq:Jgauss} (dashed dotted
line) lies on top of the estimated mean.  The dotted line is marking
the results for a Poisson process.
The sequence of solid lines in plot~(d) are the $J(r)$ function for
line--segment processes with $\varrho=100,150,200,300,400,500,600$,
bending down successively.}
\end{figure}

In Fig.~\ref{fig:FGJ-2d} the results for $F(r)$, $G(r)$, and $J(r)$,
estimated from several line--segment processes, and the Gauss--Poisson
process are shown; all the processes investigated had the same
two--point correlation function $\xi_2(r)$ given in
Eq.~\eqref{eq:xi2-line-seg}.
The line--segment process allows for larger voids than the
Gauss--Poisson process, as seen from $F_{\rm line}>F_{\rm GP}$.
On small scales the $J(r)$ of the line--segment process is well
approximated by the $J(r)$ for the Gauss--Poisson process.  However on
large scales the Gauss--Poisson process shows significantly smaller
$J(r)$ function than the line--segment process.
The $J(r)$ function is known analytically for several point process
models {}\cite{vanlieshout:j,kerscher:global}.  In any of these cases
a smaller $J(r)$ is an indication for stronger (positive) interaction
between the points (see also
{}\cite{thoennes:comparative,baddeley:estimating}).  Specifically for
Gibbs--processes (see e.g.~\cite{stoyan:stochgeom}) an attractive
interaction leads to a monotonically decreasing $J(r)$ and a stronger
interaction leads to smaller values of $J(r)$. Hence, the presence of
higher--order correlation functions in the line--segment process gives
rise to a reduced clustering strength, in the sense discussed above.
Clearly, the signal of $J(r)$ also depends on the number density.

\subsection{The non--Gaussian galaxy distribution}
\label{sect:observed-galaxy}

As already mentioned, the three--dimensional distribution of galaxies
cannot be modeled in terms of a Gauss--Poisson process: the
constraints on the density and two--point correlation function are
violated.  In the following this is illustrated with a volume--limited
sample of 100$h^{-1}$Mpc depth, extracted from the PSCz galaxy catalogue
{}\cite{saunders:pscz}.  The volume--limited sample incorporates 2232
galaxies with galactic latitude $|b|>5^\circ$. A detailed description
of the sample considered here may be found in {}\cite{kerscher:pscz}.
Estimators for the two--point correlation function are quite abundant
(see {}\cite{kerscher:comparison} and references therein).  The
results presented here do neither depend on the estimator, nor on the
exact sample geometry, which is indeed more complicated (see
{}\cite{saunders:pscz}).  For the $J(r)$--function the minus estimator
is used {}\cite{stoyan:stochgeom,kerscher:fluctuations}.

\begin{figure}
\begin{center}
\epsfig{file=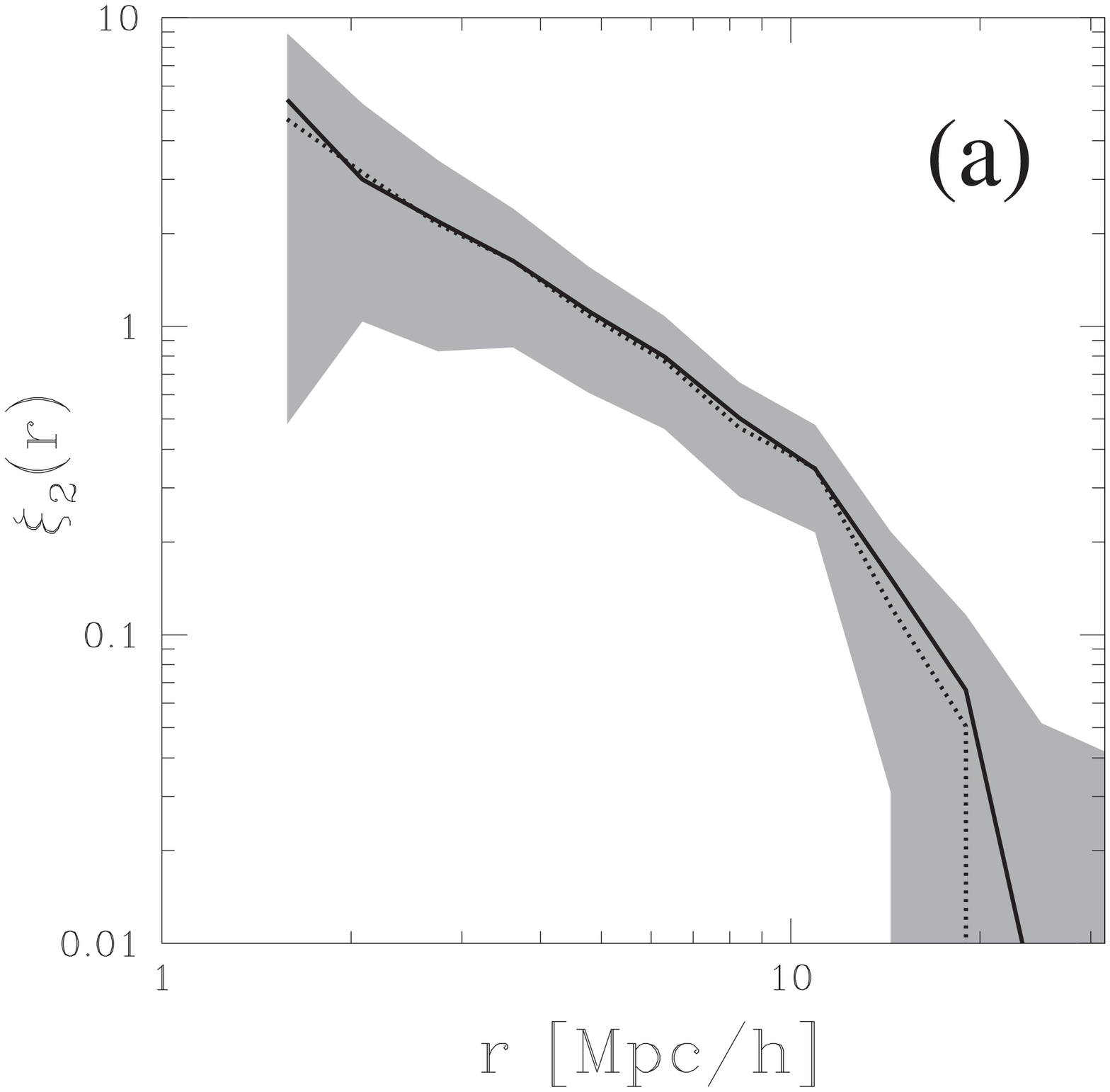,width=4.2cm}
\epsfig{file=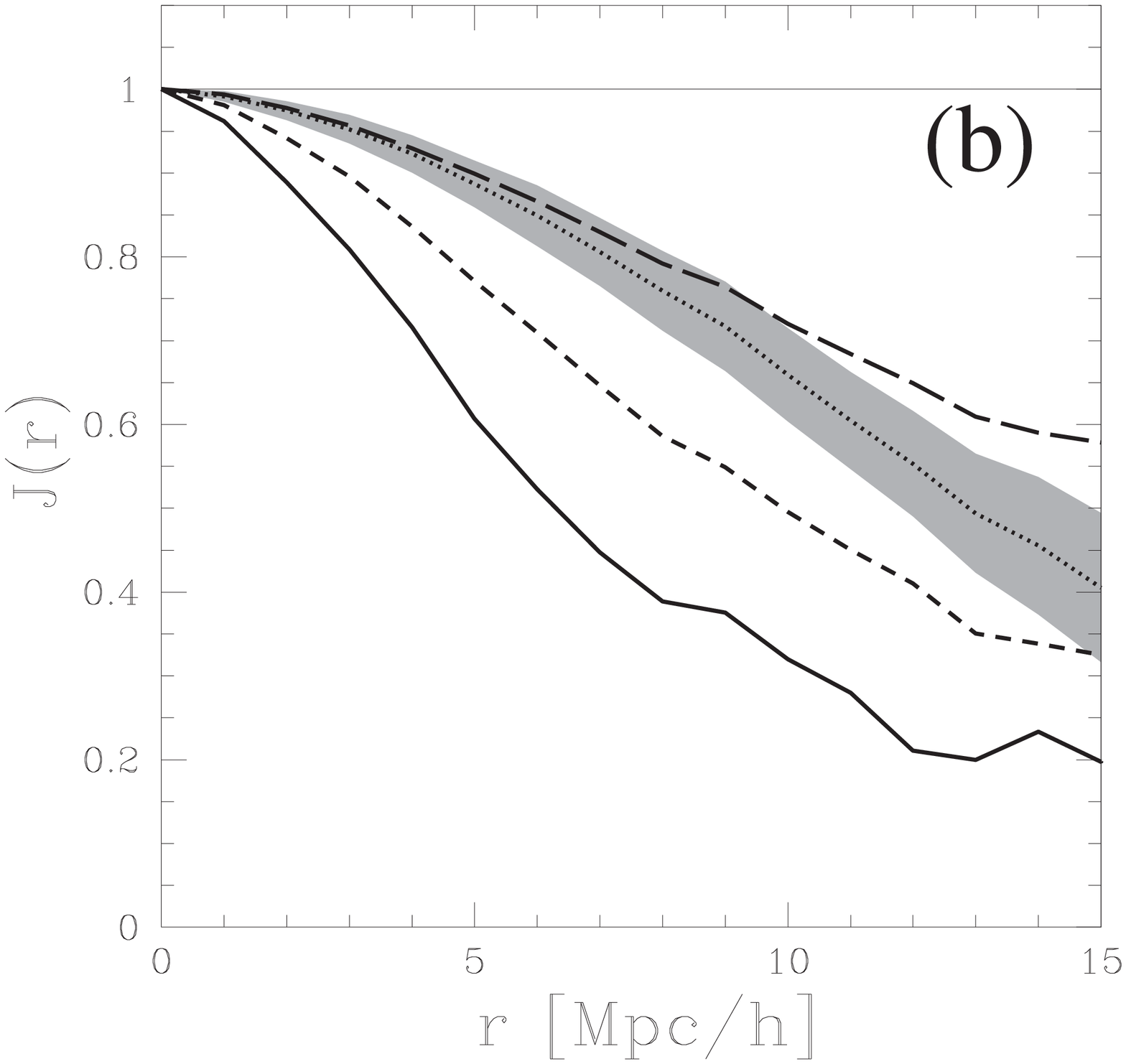,width=4.2cm}
\end{center}
\caption{\label{fig:xipscz} In plot~(a) the observed two--point
correlation function $\xi_2(r)$ of the volume--limited subsample with
100$h^{-1}$Mpc depth from the PSCz galaxy catalogue is shown (solid line).
The dotted line and the one--$\sigma$ area are estimated from 200
realizations of a Gauss--Poisson process using the estimated
two--point correlation function as an input but with only a fifth of
the number of points.
In plot~(b) the $J(r)$ function of the same sample is shown with
100\% (solid line), 50\% (short dashed line), and 20\% of the galaxies
(long dashed line).  The shaded area is the one--$\sigma$ region
obtained from the Gauss--Poisson process corresponding to the galaxy
sample with only 20\% of the points.}
\end{figure}

In Fig.~\ref{fig:xipscz} the estimated two--point correlation function
is shown. The integral
\begin{equation}
\label{eq:constraint-pscz}
\varrho\ \int_{{\mathbb{R}}^3}{\rm d}{\mathbf{y}}\ \xi_2(|{\mathbf{y}}|)\approx4.4 
\end{equation}
is violating the constraint~\eqref{eq:const2-alternate}, and the
corresponding Gauss-Poisson process does not exist.  Indeed
higher--order correlations functions have been detected by
{}\cite{szapudi:correlationspscz} using factorial moments.  By
thinning the galaxy distribution (i.e.\ randomly sub--sampling), one
generates a point set with the same correlation functions $\xi_n$ as
the observed galaxy distribution, however with a reduced number of
points.  Since the number density enters linearly in the
constraint~\eqref{eq:const2-alternate}, a comparison of the thinned
galaxy distribution with a Gauss--Poisson process becomes feasible.
The  strongly interacting  galaxy  distribution, as  indicated by  the
small values of $J(r)$,  shows increasingly weaker interaction (higher
values of $J(r)$)  for the diluted subsamples (Fig.~\ref{fig:xipscz}).

Now consider a sample with only 20\% of the actual observed galaxies,
where the constraint~\eqref{eq:const2-alternate} is satisfied (compare
with {}\eqref{eq:constraint-pscz}).  This diluted sample is
compared to a Gauss--Poisson process with the same two--point
correlation function. The $\xi_2(r)$ determined from the simulated
Gauss--Poisson process matches perfectly with the observed correlation
function (Fig.~\ref{fig:xipscz}). On small scales the $J(r)$-function
of the thinned PSCz is reasonably modeled by the Gauss--Poisson
process.  However, on large scales the Gauss--Poisson process shows
stronger interactions, whereas the thinned galaxy sample, with its
higher--order correlation functions, shows weaker interactions in the
sense discussed in Sect.~\ref{sect:detect-highorder}.

\section{Point processes with higher--order clustering}
\label{sect:higher-order}

As already mentioned, the measured two--point correlation function of
the galaxy distribution together with the observed density of galaxies
violates the constraints Eqs.~\eqref{eq:const1} and
{}\eqref{eq:const2}.  Consequently the distribution of galaxies cannot
be modeled with a Gauss--Poisson process.  Even more compelling, there
is a clear detection of higher--order correlations in the galaxy
distribution (e.g.\
{}\cite{fry:statistical,bonometto:correlation,szapudi:comparison}).
Hence, one is interested in analytical tractable approximations of the
cumulant expansion~\eqref{eq:cumulant-exp}.  Hierarchical closure
relations have been extensively studied (see
Sect.~\ref{sect:hierarchical}).
In the following a truncation of the expansion~\eqref{eq:cumulant-exp}
beyond the Gaussian term and the $n$--point Poisson cluster processes
will be used.

Such a truncation may serve as a model for the galaxy distribution in
the weakly nonlinear regime.  Using perturbation theory one may show
that $\xi_3\propto(\xi_2)^2$ (see {}\cite{fry:galaxy}). For large
separations $r$ the correlation function $\xi_2(r)$ is smaller than
unity, and consequently a truncation of the
expansion~\eqref{eq:cumulant-exp} at $n>2$ provides a viable model for
the large--scale distribution of galaxies.

The general Poisson cluster process is the starting point: consider
the expansion of the cluster p.g.fl.\ $G_c[h|{\mathbf{x}}]$ in terms
of Janossy densities conditional on the cluster center ${\mathbf{x}}$
(see Eq.~\eqref{eq:janossy-exp}):
\begin{multline}
\label{eq:cluster-exp-janossy}
G_c[h|{\mathbf{x}}] = J_0 + \sum_{n=1}^\infty \frac{1}{n!}
\int_{{\mathbb{R}}^d}{\rm d}{\mathbf{x}}_1\cdots\int_{{\mathbb{R}}^d}{\rm d}{\mathbf{x}}_n \\
j_n({\mathbf{x}}_1,\ldots{\mathbf{x}}_n|{\mathbf{x}})\ h({\mathbf{x}}_1)\cdots h({\mathbf{x}}_n).
\end{multline}
Explicit expression for the Janossy densities are given below. 
The p.g.fl.\ of a Poisson cluster process is then given by
\begin{align}
\label{eq:general--poisson-cluster}
G[h] & = \exp\Bigg( \int_{{\mathbb{R}}^d}{\rm d}{\mathbf{x}}\varrho_c\ (G_c[h|{\mathbf{x}}]-1)\Bigg)
\nonumber\\
& =\exp\Bigg( \int_{{\mathbb{R}}^d}{\rm d}{\mathbf{x}}\varrho_c\ 
\Big( \sum_{n=1}^\infty  \frac{1}{n!}
\int_{{\mathbb{R}}^d}{\rm d}{\mathbf{x}}_1\cdots\int_{{\mathbb{R}}^d}{\rm d}{\mathbf{x}}_n 
\nonumber\\
& \qquad
j_n({\mathbf{x}}_1,\ldots{\mathbf{x}}_n|{\mathbf{x}})\ h({\mathbf{x}}_1)\cdots h({\mathbf{x}}_n) -1 \Big) \Bigg) .
\end{align}
Here the probability $q_0$ of having no point in the cluster at ${\mathbf{x}}$ 
is assumed to be zero, i.e.\ $J_0=0$.  This does not impose any 
additional constraints, it only leads to a redefinition of the number 
density of cluster centers $\varrho_c'=\varrho_c(1+q_0)$.

Using  this more formal  approach the  p.g.fl.\ of  the Gauss--Poisson
process can be written in  terms of the Janossy densities with $j_n=0$
for $n>2$:
\begin{equation}
\label{eq:cluster-janossy-2}
\begin{split}
j_1({\mathbf{x}}_1|{\mathbf{x}}) & = q_1({\mathbf{x}})\delta^D({\mathbf{x}}-{\mathbf{x}}_1), \\
j_2({\mathbf{x}}_1,{\mathbf{x}}_2|{\mathbf{x}}) &  = q_2({\mathbf{x}}) \delta^D({\mathbf{x}}-{\mathbf{x}}_1) f_2({\mathbf{x}}_2|{\mathbf{x}}_1)\ 2! .
\end{split}
\end{equation}
Here $j_1$ ($j_2$) are the probability densities for the spatial
distribution of one (two) points in the cluster, multiplied by the
probability $q_1$ ($q_2$) that there are exactly one (two) points in
the cluster at ${\mathbf{x}}$.  $\delta^D$ is the $d$--dimensional Dirac
distribution.  $f_2({\mathbf{x}}_2|{\mathbf{x}}_1)$ is the probability density of the
second point ${\mathbf{x}}_2$ under the condition that there is a point at
${\mathbf{x}}_1$, normalized by $\int{\rm d}{\mathbf{x}}_2\ f_2({\mathbf{x}}_2|{\mathbf{x}}_1)=1$.

The p.g.fl.\ {}\eqref{eq:general--poisson-cluster} is invariant under
changes of the order of integration, implying that one can use the
$j_n({\mathbf{x}}_1,\ldots{\mathbf{x}}_n|{\mathbf{x}})$ symmetrically defined in all coordinates
(including ${\mathbf{x}}$).  With the additional assumption of homogeneity and
isotropy one gets $f_2({\mathbf{x}}_2|{\mathbf{x}}_1)=f(|{\mathbf{x}}_1-{\mathbf{x}}_2|)$, as already used in
Sect.~\ref{sect:gp-as-pc} for the construction of the Gauss--Poisson
process.

\subsection{The three--point Poisson cluster process}
\label{sect:three-point}

In a three--point Poisson cluster process
Eq.~\eqref{eq:general--poisson-cluster} is truncated at the third
order and at most three points per cluster are allowed.  Additional to
Eq.~\eqref{eq:cluster-janossy-2}
\begin{equation}
j_3({\mathbf{x}}_1,{\mathbf{x}}_2,{\mathbf{x}}_3|{\mathbf{x}}) = 
q_3({\mathbf{x}}) \delta^D({\mathbf{x}}-{\mathbf{x}}_1) f_3({\mathbf{x}}_2,{\mathbf{x}}_3|{\mathbf{x}}_1)\ 3! 
\end{equation}
appears, with the probability $q_3({\mathbf{x}})$ that the cluster consists out
of three points, and $q_1+q_2+q_3=1$ with $q_i\ge0$.
$f_3({\mathbf{x}}_2,{\mathbf{x}}_3|{\mathbf{x}}_1)$ is the probability density that there are two
points at ${\mathbf{x}}_2$, and ${\mathbf{x}}_3$, under the condition that one point is
at ${\mathbf{x}}_1$, with the normalization $\int{\rm d}{\mathbf{x}}_2\int{\rm d}{\mathbf{x}}_3\
f_3({\mathbf{x}}_2,{\mathbf{x}}_3|{\mathbf{x}}_1)=1$.
Inserting these definitions one obtains
\begin{multline}
\log G[h] = \int_{{\mathbb{R}}^d}{\rm d}{\mathbf{x}}_1\ \varrho_p q_1({\mathbf{x}}_1)\ (h({\mathbf{x}}_1)-1) +\\
+ \int_{{\mathbb{R}}^d}{\rm d}{\mathbf{x}}_1\ \varrho_p \int_{{\mathbb{R}}^d}{\rm d}{\mathbf{x}}_2\ 
q_2({\mathbf{x}}_1) f_2({\mathbf{x}}_2|{\mathbf{x}}_1)\ (h({\mathbf{x}}_1)h({\mathbf{x}}_2)-1) +\\
+ \int_{{\mathbb{R}}^d}{\rm d}{\mathbf{x}}_1\ \varrho_p \int_{{\mathbb{R}}^d}{\rm d}{\mathbf{x}}_2\int_{{\mathbb{R}}^d}{\rm d}{\mathbf{x}}_3\ 
q_3({\mathbf{x}}_1) f_3({\mathbf{x}}_2,{\mathbf{x}}_3|{\mathbf{x}}_1)\times\\ 
\times\Big(h({\mathbf{x}}_1)h({\mathbf{x}}_2)h({\mathbf{x}}_3)-1\Big) .
\end{multline}
As already mentioned, $f_3({\mathbf{x}}_2,{\mathbf{x}}_3|{\mathbf{x}}_1)$ can be assumed to be
symmetric in its three arguments.  Slightly abusing notation, let
$f_2({\mathbf{x}}_1,{\mathbf{x}}_2)$ and $f_3({\mathbf{x}}_1,{\mathbf{x}}_2,{\mathbf{x}}_3)$ be the symmetrically
defined densities corresponding to $f_2({\mathbf{x}}_2|{\mathbf{x}}_1)$ and
$f_3({\mathbf{x}}_2,{\mathbf{x}}_3|{\mathbf{x}}_1)$, and define
\begin{equation}
\label{eq:def-g2}
f_2^{(3)}({\mathbf{x}}_1,{\mathbf{x}}_2) = \int_{{\mathbb{R}}^d}{\rm d}{\mathbf{x}}_3\ f_3({\mathbf{x}}_1,{\mathbf{x}}_2,{\mathbf{x}}_3).
\end{equation}
Replacing  $h$  by  $h+1$  and  rearranging the  terms  the  factorial
cumulant expansion of the three--point cluster process reads
\begin{multline}
\label{eq:pgfl-cumulant-three}
\log G[h+1] =  \int_{{\mathbb{R}}^d}{\rm d}{\mathbf{x}}_1\ h({\mathbf{x}}_1)\
\varrho_p \big(1+q_2({\mathbf{x}}_1)+2q_3({\mathbf{x}}_1)\big) + \\
\begin{aligned}
+& \int_{{\mathbb{R}}^d}{\rm d}{\mathbf{x}}_1\int_{{\mathbb{R}}^d}{\rm d}{\mathbf{x}}_2\ h({\mathbf{x}}_1)h({\mathbf{x}}_2)\times\\ 
&\quad \times\varrho_p 
\Big(q_2({\mathbf{x}}_1) f_2({\mathbf{x}}_1,{\mathbf{x}}_2) + 3q_3({\mathbf{x}}_1) f_2^{(3)}({\mathbf{x}}_1,{\mathbf{x}}_2)\Big) +\\
+&\int_{{\mathbb{R}}^d}{\rm d}{\mathbf{x}}_1\int_{{\mathbb{R}}^d}{\rm d}{\mathbf{x}}_2\int_{{\mathbb{R}}^d}{\rm d}{\mathbf{x}}_3\  
h({\mathbf{x}}_1)h({\mathbf{x}}_2)h({\mathbf{x}}_3)\times\\ 
&\quad \times\varrho_p q_3({\mathbf{x}}_1) f_3({\mathbf{x}}_1,{\mathbf{x}}_2,{\mathbf{x}}_3) . 
\end{aligned}
\end{multline}
Comparing Eq.~\eqref{eq:pgfl-cumulant-three} with the
expansion~\eqref{eq:cumulant-exp} one arrives at
\begin{align}
\label{eq:f-xi-three}
\varrho & = (q_1+2q_2+3q_3) \varrho_p = (1+q_2+2q_3) \varrho_p \nonumber\\
\xi_2({\mathbf{x}}_1,{\mathbf{x}}_2) 
& = \frac{2!}{(1+q_2+2q_3)^2 \varrho_p}\times\nonumber\\
& \quad \times\Big(q_2 f_2({\mathbf{x}}_1,{\mathbf{x}}_2) + 3q_3 f_2^{(3)}({\mathbf{x}}_1,{\mathbf{x}}_2)\Big)
\nonumber\\
\xi_3({\mathbf{x}}_1,{\mathbf{x}}_2,{\mathbf{x}}_3) 
& = \frac{3!}{(1+q_2+2q_3)^3 \varrho_p^2}\  q_3 f_3({\mathbf{x}}_1,{\mathbf{x}}_2,{\mathbf{x}}_3),
\end{align}
and the correlation functions $\xi_n$ equal zero for $n\ge4$.
The simulation procedure for the three--point Poisson cluster process
is described in Appendix~\ref{sect:simul-threepoint}.

\subsection{Constraints on $\varrho$, $\xi_2$ and $\xi_3$}
\label{sect:const-three}

By the definition of the three--point Poisson cluster process, the
probability densities $f_2\ge0$, $f_3\ge0$, and $f_2^{(3)}\ge0$ and
consequently $\xi_2(r)\ge0$ for all $r$, as well as $\xi_3\ge0$.
This is a generic feature of Poisson cluster processes.

The Gauss--Poisson process, defined through the truncation of the 
cumulant expansion after the second term, is equivalent to the 
two--point Poisson cluster process (see Sect.~\ref{sect:gp-as-pc}).  
Unfortunately, this equivalence does not hold for the higher $n$--point 
processes anymore.
The general three--point process is defined as point process with a
factorial cumulant expansion truncated after the third term.
Proceeding similar to Sect.~\ref{sect:constraints-gauss},
necessary conditions for the existence of such a point process can be
derived (compare with Eq.~\eqref{eq:const-general}):
\begin{multline}
0\le \varrho|A_l|
+\sum_{i=1}^k \varrho^2|A_l||A_i|\overline{\xi}_2(A_l,A_i)(z_i-1) +\\
+\sum_{i=1}^k\sum_{j=1}^k \frac{\varrho^3}{2}|A_l||A_i||A_j|
\overline{\xi}_3(A_l,A_i,A_j)(z_i-1)(z_j-1) ,
\end{multline}
with the volume--averaged correlation functions
\begin{multline}
\label{eq:def-volume-averaged-xi}
\overline{\xi}_n(A_1,\ldots,A_n) = \frac{1}{|A_1|\cdots|A_n|} \times\\ 
\times\int_{A_1}{\rm d}{\mathbf{x}}_1\cdots\int_{A_n}{\rm d}{\mathbf{x}}_n\
\xi_n({\mathbf{x}}_1,\ldots,{\mathbf{x}}_n) ,
\end{multline}
and for consistency $\overline{\xi}_1(A)=1$.
Again, for $z_i=1$ one obtains $\varrho\ge0$. The non--trivial
constraints read:
\begin{align}
\label{eq:const3p-1}
1 \ge &\ \varrho|A_s| \overline{\xi}_2(A_l,A_s) - 
\varrho^2 |A_s|^2 \overline{\xi}_3(A_l,A_s,A_s), \\
\label{eq:const3p-2}
0 \le &\ \overline{\xi}_3(A_l,A_s,A_s) , \\
\label{eq:const3p-3}
1\ge &\ \varrho|A_s| \overline{\xi}_2(A_l,A_s) + 
\varrho|A_r|\overline{\xi}_2(A_l,A_r) - \nonumber\\
&\ - \frac{\varrho^2}{2} |A_s|^2 \overline{\xi}_3(A_l,A_s,A_s) 
- \frac{\varrho^2}{2} |A_r|^2 \overline{\xi}_3(A_l,A_r,A_r) -
\nonumber\\
&\ - \varrho^2 |A_s| |A_r| \overline{\xi}_3(A_l,A_s,A_r) .
\end{align}
Eq.~\eqref{eq:const3p-1} can be derived by setting $z_s=0$ and $z_i=1$
for all $i\ne s$, Eq.~\eqref{eq:const3p-2} follows from
$z_s\rightarrow\infty$ and $z_i=1$ for all $i\ne s$.  Using
$z_r,z_s\rightarrow\infty$ does not lead to new constraints.  With
$z_r=0=z_s$, $r\ne s$, and $z_i=1$ for all $i\ne r,s$ one obtains
Eq.~\eqref{eq:const3p-3}.  No additional constraint arises by setting
$z=2$.

Eq.~\eqref{eq:const3p-2} implies $\xi_3\ge0$.
Eq.~\eqref{eq:const3p-1} and Eq.~\eqref{eq:const3p-3} are the
extension of the constraint~\eqref{eq:const1a}.  The terms
proportional to $\overline{\xi}_3$ can balance the terms with
$\overline{\xi}_2$, and a clustering point processes with a number
density higher than in a Gauss--Poisson process is possible.
Moreover, $\xi_2$ is not constrained to positive values anymore.
Hence, already by including three--point correlations, a point process
model with a two--point correlation function $\xi_2$ having a zero
crossing becomes admissible.  This answers affirmatively the question
by {}\cite{milne:generalized}, whether there exists a general
three--point cluster processes with a negative second moment.
However, in the three--point Poisson cluster process discussed in the
preceding section a $\xi_2\ge0$ is required illustrating that the
three--point Poisson cluster processes form only a subset of all
possible three--point processes.

\subsection{The $n$--point Poisson cluster process}

It  is now  clear  how  to construct  the  $n$--point Poisson  cluster
process.   Let $q_m$  be  the  probability of  having  $m$ points  per
cluster with $\sum_{m=1}^nq_m=1$.
\begin{equation}
\label{eq:def-jn}
j_n({\mathbf{x}}_1,\ldots,{\mathbf{x}}_n|{\mathbf{x}}) = 
q_n({\mathbf{x}}) \delta^D({\mathbf{x}}-{\mathbf{x}}_1) f_n({\mathbf{x}}_2,\ldots,{\mathbf{x}}_n|{\mathbf{x}}_1)\ n! ,
\end{equation}
determines the distribution of the $n$ points inside the cluster 
($f_1=1$).  As above the $f_n$ are assumed to be symmetric in all 
their arguments, and for $n>m$
\begin{multline}
\label{eq:def-gnm}
f_m^{(n)}({\mathbf{x}}_1,\ldots,{\mathbf{x}}_m) = 
\int_{{\mathbb{R}}^d}{\rm d}{\mathbf{x}}_{m+1}\cdots\int_{{\mathbb{R}}^d}{\rm d}{\mathbf{x}}_n\\
f_n({\mathbf{x}}_1,\ldots,{\mathbf{x}}_m,{\mathbf{x}}_{m+1},\ldots,{\mathbf{x}}_n),
\end{multline}
and $f_m=f_m^{(m)}$.
Inserting Eq.~\eqref{eq:def-jn} into
Eq.~\eqref{eq:general--poisson-cluster} and after some algebraic
manipulations one can compare term by term with the factorial cumulant
expansion~\eqref{eq:cumulant-exp} of the p.g.fl.:
\begin{align}
\label{eq:moments-n-point-poi}
\varrho & =\varrho_p\sum_{m=1}^n m q_m \\
\xi_k({\mathbf{x}}_1,\ldots,{\mathbf{x}}_k) 
&= \frac{k! \varrho_p}{\varrho^k} 
\sum_{m=k}^n \binom{m}{k} q_m f_k^{(m)}({\mathbf{x}}_1,\ldots,{\mathbf{x}}_k),\nonumber
\end{align}
with $k\le n$, and $\xi_k=0$ for $k>n$.  The statistical properties of 
this $n$--point Poisson cluster process are now completely specified 
by the correlation functions $\xi_k$ with $k\le n$ and the mean 
density $\varrho$.
Eqs.~\eqref{eq:moments-n-point-poi} and the normalization of the $f_m$
can be used to determine the $f_m$ as well as $\varrho_p$ and $q_m$
from given correlation functions $\xi_n$ and the number density
$\varrho$.  A simulation algorithm similar to the one described in
Appendix~\ref{sect:simul-threepoint} can be constructed.

\subsection{The general $n$--point process}
\label{sect:general-n-point}

The general $n$--point process is defined as the point process
resulting from the factorial cumulant expansion truncated after the
$n$th term.  Proceeding similarly to
Sect.~\ref{sect:constraints-gauss} one arrives at the constraint
equations
\begin{multline}
0\ge \varrho|A_l| 
+\varrho^2\sum_{i=1}^k |A_l||A_i|
\overline{\xi}_2(A_l,A_i)(z_i-1) +\ldots+ \\
+\frac{\varrho^n n}{n!}\sum_{i_1,\ldots,i_{n-1}=1}^k 
|A_l||A_{i_1}|\cdots|A_{i_{n-1}}|\times\\
\times\overline{\xi}_n(A_l,A_{i_1},\ldots,A_{i_{n-1}})
(z_{i_1}-1)\cdots(z_{i_{n-1}}-1) .
\end{multline}
It is now possible to compute the constraints for the $n$--point
process, in close analogy to the three--point process in
Sect.~\ref{sect:const-three}.  {}\cite{milne:generalized} gave
necessary and sufficient conditions for the existence of a generalized
Hermite distribution (closely related to this $n$--point process).
They discuss the constraints for a slightly different expansion of the
p.g.f.  Unfortunately, the transformation of their expansion to the
expansion in terms of correlation functions is as tedious as the
direct calculation of the constraints.

\section{Random fields vs. point processes}
\label{sect:field-point}

A random field $u({\mathbf{x}})$ is in the simplest case a real--valued
function on ${\mathbb{R}}^d$ {}\cite{adler:randomfields}. In cosmology the
initial mass--density field is often modeled as a Gaussian random
field (see e.g.~\cite{bardeen:gauss,sahni:approximation}). The
nonlinear evolution of the density field unavoidably introduces
higher--order correlations.  A random field $u(x)$ is
stochastically characterized by its characteristic functional
(e.g.~\cite{vankampen:stochastic})
\begin{equation}
\Phi^u[v]={\mathbb{E}}^u\Bigg[\exp\Bigg(i\int_{{\mathbb{R}}^d}{\rm d}{\mathbf{x}}\ v({\mathbf{x}})u({\mathbf{x}})\Bigg)\Bigg],
\end{equation}
where ${\mathbb{E}}^u$ denotes the expectation value over realizations of the
random field $u$.  In close analogy to the
expansion~\eqref{eq:phi-moment-expansion} of the characteristic
function of a random variable in terms of cumulants, one obtains the
expansion of the characteristic functional
\begin{multline}
\ln\Phi^u[v] = \sum_{n=1}^\infty \frac{i^n}{n!}
\int_{{\mathbb{R}}^d}{\rm d}{\mathbf{x}}_1\cdots\int_{{\mathbb{R}}^d}{\rm d}{\mathbf{x}}_n \\
c_n^u({\mathbf{x}}_1,\ldots,{\mathbf{x}}_n)\ v({\mathbf{x}}_1)\cdots v({\mathbf{x}}_n) 
\end{multline}
in terms of $n$-point cumulants
$c_n^u({\mathbf{x}}_1,\ldots,{\mathbf{x}}_n)$. Here
$c_1^u({\mathbf{x}})={\mathbb{E}}^u[u({\mathbf{x}})]=\overline{u}$ is the mean value. The
correlation function of the field is
\begin{equation}
\xi_2^u({\mathbf{x}}_1,{\mathbf{x}}_2) 
= \frac{c_2^u({\mathbf{x}}_1,{\mathbf{x}}_2)}{\overline{u}^2}
= \frac{{\mathbb{E}}^u[u({\mathbf{x}}_1)u({\mathbf{x}}_2)]}{\overline{u}^2}-1,
\end{equation}
and similar for higher--order correlation functions (see
e.g.~\cite{vankampen:stochastic,borgani:scaling}).  The well known
characteristic functional of the Gaussian random field reads
\begin{multline}
\ln\Phi^u[v] = i\int_{{\mathbb{R}}^d}{\rm d}{\mathbf{x}}_1\
\overline{u} v({\mathbf{x}}_1) - \\
- \frac{1}{2}\int_{{\mathbb{R}}^d}{\rm d}{\mathbf{x}}_1\int_{{\mathbb{R}}^d}{\rm d}{\mathbf{x}}_2\ 
c_2^u({\mathbf{x}}_1,{\mathbf{x}}_2)\ v({\mathbf{x}}_1)v({\mathbf{x}}_2),
\end{multline}
with the covariance function $c_2^u({\mathbf{x}}_1,{\mathbf{x}}_2)$.

The characteristic functional of a point process is defined by
\begin{equation}
\Phi[h] = {\mathbb{E}}\left[
\exp\left(i\int_{{\mathbb{R}}^d}N({\rm d} {\mathbf{x}}) h({\mathbf{x}}) \right) \right] ,
\end{equation}
and the relation to the p.g.fl.\ is  $\Phi[h]=G[{\rm e}^{ih}]$. An
expansion into cumulants  $c_k({\mathbf{x}}_1,...,{\mathbf{x}}_k)$ is also possible:
\begin{multline}
\log\Phi[h] = \sum_{k=1}^\infty \frac{i^k}{k!} 
\int_{{\mathbb{R}}^d}{\rm d}{\mathbf{x}}_1\cdots\int_{{\mathbb{R}}^d}{\rm d}{\mathbf{x}}_k \\
c_k({\mathbf{x}}_1,\ldots,{\mathbf{x}}_k)\ h({\mathbf{x}}_1)\cdots h({\mathbf{x}}_k).
\end{multline}
The cumulants $c_k(\cdot)$ should not be confused with the factorial
cumulants $c_{[k]}(\cdot)$.

\subsection{A theorem of Marcinkiewicz}
\label{sect:marcinkiewicz}

A theorem of Marcinkiewicz {}\cite{marcinkiewicz:sur} states that if
the characteristic function $\varphi(s)$ of a random variable (see
Appendix~\ref{sect:random-variables}) is the exponential of a
polynomial with finite degree larger than two, then the positive
definiteness of the probability distribution is violated (see e.g.\
{}\cite{richter:wahrscheinlichkeitstheorie,pawula:approximation,linnik:decomposition}).
The generalized Marcinkiewicz theorem for characteristic functionals
{}\cite{robinson:theorem,rajagopal:generalization} tells us that this
expansion has to be either infinite or a polynomial in $h({\mathbf{x}})$ (or
$v({\mathbf{x}})$) of degree less than or equal to two.
This directly applies to the expansion of the characteristic
functionals of a random field $\Phi^u[v]$ and a point process
$\Phi[h]$ in terms of cumulants.

However, for a point process one can see that the expansion of the
p.g.fl.\ in terms of {\em factorial} cumulants $c_{[k]}(\cdot)$ (or
correlation functions $\xi_k(\cdot)$) allows a truncation at a finite
$k>2$. As long as constraints on the density and the correlation
functions are fulfilled, the point process is well defined.
Although the p.g.fl.\ was used mainly in the context of discrete
events, it seems worthwhile to consider the characterizations of
random fields with factorial cumulants.

Another systematic expansion is provided by the Edgeworth series.  It
was successfully applied in cosmology to quantify the one point
probability distribution function for the smoothed density field on
large scales {}\cite{juszkiewicz:weakly}.  Recently,
{}\cite{contaldi:generating} showed how to use the truncated Edgeworth
series to generate realizations of non--Gaussian random fields with
predefined correlation properties. The truncated Edgeworth series also
violates the positive definiteness of the probability distribution,
but {}\cite{contaldi:generating} restore the positivity, reintroducing
higher correlations, leading to a ``leaking'' into higher
correlations.

The cumulants $c_k$ and the factorial cumulants $c_{[k]}$ of a random
variable are related by $c_k=\sum_{l=1}^k s_2(k,l)\ c_{[l]}$ (see
Eq.~\eqref{eq:rv-cumulant-fact-cumulant}). Looking at the
Poisson cluster processes discussed in the preceding sections, one
observes that such a relation must not hold between the cumulants and
the factorial cumulants of a point process.
As an example consider the three--point Poisson cluster process with
$c_{[n]}(\cdot)=0$ for all $n>3$. A finite $c_{[3]}(\cdot)$ leads to
non--zero $c_{n}(\cdot)$ for all $n$ (see
Appendix~\ref{sect:cumulants-factorial-cumulants} for details.)

\subsection{The Poisson model}

In cosmology the point distribution is often related to the mass
density field assuming the ``Poisson model''. The value of the mass
density field is assumed to be proportional to the local number
density, and the point distribution is constructed by ``Poisson
sampling'' the correlated mass density field. If the mass density
field is itself a realization of a random field, the resulting point
process is called a Cox process, or doubly stochastic process.
Within this model one may show that the cumulants $c_n^u$ of the
density field are proportional to the {\em factorial} cumulants
$c_{[n]}$ of the point distribution
{}\cite{peebles:lss,daley:introduction}: $c_{[n]}=c_n^u\
\rho^2/\overline{u}^2$. It is important to notice that this relates
the characteristic functional of the random field $\Phi^u[v]$ with the
p.g.fl\ of the point process $G[h]$.
For the correlation functions one obtains
\begin{equation}
\xi_n({\mathbf{x}}_1,\ldots,{\mathbf{x}}_n) = \xi_n^u({\mathbf{x}}_1,\ldots,{\mathbf{x}}_n) .
\end{equation}
Hence, this model allows the direct comparison of predictions from
analytical calculations with the observed correlation functions in the
galaxy distribution.

Clearly the question arises, what is wrong with the simple picture
that one starts with a Gaussian random field and ``Poisson sample'' it
to obtain the desired point distribution.  The answer is that a
Gaussian random field is an approximate model for a mass density field
only if the fluctuations are significantly smaller than the mean mass
density.  Otherwise negative mass densities (i.e.\ negative
``probabilities'' for the Poisson sampling) would occur. Only in the
limit of vanishing fluctuations a Poisson sampled Gaussian random
field becomes a permissible model. However, in this limit one is left
with a pure Poisson process.

\section{Models for strongly correlated systems}
\label{sect:strongly}

In the Sects.~\ref{sect:gauss-poisson} and {}\ref{sect:higher-order}
several types of point processes were discussed, all featuring a
truncated factorial cumulant expansion. As argued at the beginning of
Sect.~\ref{sect:higher-order}, such a truncation is feasible for the
matter distribution in the Universe, as long as $\xi_2(r)<1$, i.e.\
for points with large separations.
Mainly two approaches have been followed to model the galaxy
distribution also on small scales with $\xi_2(r)\gg1$. The
hierarchical models are briefly discussed in the next section and in
Sect.~\ref{sect:beyond-halo} an extension of the halo--model is
presented.

\subsection{Hierarchical models}
\label{sect:hierarchical}

In cosmology one often starts with a scale--invariant correlation
function $\xi_2(r)\propto{r}^{-\gamma}$ and assumes some closure
relations for the $\xi_n$.  Especially the hierarchical ansatz
$\xi_n=Q_n\sum_{\text{trees}}\prod^{n-1}\xi_2$ was extensively studied
(e.g.\
{}\cite{peebles:lss,fry:four,fry:galaxy-npoint,balian:I,caruthers:galaxy,szapudi:higher},
and more recent {}\cite{scoccimarro:hyperextended,bernardeau:halo}).
{}\cite{balian:I} discuss conditions for the coefficients $Q_n$ such
that the expansion of the p.g.f.'s\ in terms of the count--in--cells
converges.  In this case the count--in--cells uniquely determine the
point process. As illustrated in Sect.~\ref{sect:problems} with the
log--normal distribution, a non--converging expansion does not
necessarily imply that the stochastic model is not well--defined.  It
only implies that such a point process model is not completely
specified by its correlation functions.  For critical systems similar
expansion in terms of correlation functions are typically divergent
(see e.g.\ {}\cite{zinnjustin:quantum}, chapt.~41).

As another closure relation Kirkwood
{}\cite{kirkwood:statistical} employed the following approximation 
\begin{multline}
\varrho_3({\mathbf{x}}_1,{\mathbf{x}}_2,{\mathbf{x}}_3)=\\
\varrho^3(1+\xi_2({\mathbf{x}}_1,{\mathbf{x}}_2))
(1+\xi_2({\mathbf{x}}_2,{\mathbf{x}}_3))(1+\xi_2({\mathbf{x}}_3,{\mathbf{x}}_1))
\end{multline}
to calculate thermodynamic properties of fluids using the BBGKY hierarchy. This closure relation is exact for the
log--normal distribution (e.g.\ {}\cite{coles:lognormal}).
Empirically however one finds that this ansatz is disfavored as a
model for the galaxy distribution {}\cite{szapudi:higherapm}.

\subsection{The generalized halo model}
\label{sect:beyond-halo}

In Sect.~\ref{sect:higher-order} several types of Poisson cluster
processes were constructed by starting with Poisson distributed
centers and attaching a secondary point process, the cluster, to each
point.  One can generalize this procedure by considering cluster
centers given by already correlated points.  One possibility, is to
iterate the construction principle of the simple Poisson cluster
process leading to the $m$--th order Neyman--Scott processes
{}\cite{neyman:statistical}.
If one is only interested in the first few correlation functions, the
full specification of the point process is not necessary.  Within the
halo model (see e.g.\
{}\cite{scherrer:statistics,sheth:non-linear,ma:halo,ma:deriving,peacock:halo,scoccimarro:howmany})
it is specifically easy to calculate the correlation functions.  The
difference to the Poisson cluster processes discussed previously
is that the cluster centers now may be correlated themselves.
The major physical assumption entering is that the properties of the
clusters (halos) are {\em independent} from the positions and
correlations of the cluster centers.

Consider a point process for the cluster centers, the parents,
specified by the p.g.fl.\ $G_p[h]$.  Independent from the distribution
of the centers, a cluster with a p.g.fl.\ $G_c[h|{\mathbf{y}}]$ is
attached to each center ${\mathbf{y}}$.  Then the p.g.fl.\ of this
cluster process is given by the ``folding'' of the two p.g.fl.'s
{}\cite{daley:introduction}:
\begin{equation}
\label{eq:general-cluster-process}
G[h]  = G_p\big[G_c[h|\cdot]\big].
\end{equation}
Using the expansion~\eqref{eq:cumulant-exp}, these p.g.fl.'s are given 
by
\begin{align}
\label{eq:general-cluster-parent}
\log G_p[h+1] &= \sum_{m=1}^\infty    
\frac{\varrho_p^m}{m!}\int_{{\mathbb{R}}^d}{\rm d}{\mathbf{y}}_1\cdots\int_{{\mathbb{R}}^d}{\rm d}{\mathbf{y}}_m 
\nonumber\\
& \quad \xi_m^{(p)}({\mathbf{y}}_1,\ldots,{\mathbf{y}}_m)\ h({\mathbf{y}}_1)\cdots h({\mathbf{y}}_m),\\
\log G_c[h+1|{\mathbf{y}}] &= \sum_{n=1}^\infty    
\frac{1}{n!}\int_{{\mathbb{R}}^d}{\rm d}{\mathbf{x}}_1\cdots\int_{{\mathbb{R}}^d}{\rm d}{\mathbf{x}}_n \nonumber\\
& \quad c_{[n]}({\mathbf{x}}_1,\ldots,{\mathbf{x}}_n|{\mathbf{y}})\ h({\mathbf{x}}_1)\cdots h({\mathbf{x}}_n),
\end{align}
where $\varrho_p$ is the number density and the $\xi_n^{(p)}$ are the
correlation functions of the parent process.  The
$c_{[n]}(\ldots|{\mathbf{y}})$ are the factorial cumulants specifying the point
distribution in a cluster, conditional on the cluster center ${\mathbf{y}}$.
$c_{[1]}({\mathbf{x}}|{\mathbf{y}})$ is the halo profile, with the mean number of points
per halo $\mu=\int{\rm d}{\mathbf{x}}\ c_{[1]}({\mathbf{x}}|{\mathbf{y}})$.  The $c_{[n]}$, $n\ge2$
quantify the halo substructure. A halo without substructure is an
inhomogeneous Poisson process, and completely characterized by
$c_{[1]}({\mathbf{x}}|{\mathbf{y}})$ and $c_{[n]}=0$, $n\ge2$.
Combining Eqs.~\eqref{eq:general-cluster-process} and
{}\eqref{eq:general-cluster-parent}
\begin{multline}
\label{eq:halo-expansion}
\log G[h+1] = 
\sum_{m=1}^\infty    
\frac{\varrho_p^m}{m!}\int_{{\mathbb{R}}^d}{\rm d}{\mathbf{y}}_1\cdots\int_{{\mathbb{R}}^d}{\rm d}{\mathbf{y}}_m \\
\xi_m^{(p)}({\mathbf{y}}_1,\ldots,{\mathbf{y}}_m)\ \prod_{i=1}^m 
\big(G_c[h+1|{\mathbf{y}}_i]-1\big) , 
\end{multline}
one immediately recovers the p.g.fl.\ of the Poisson cluster process 
Eq.~\eqref{eq:poisson-cluster-G} by setting $\xi_m^{(p)}=0$ for 
$m\ge2$ ($\xi_1^{(p)}=1$).

\subsection{$\xi_2$ and $\xi_3$ in the generalized halo model}

In the standard halo model the clusters are simply modeled by an
inhomogeneous Poisson process, whereas the centers are given by a
correlated point process, typically determined from the evolved
density distribution.  Based on these assumptions one can calculate
the correlation functions $\xi_n$ for the halo model
{}\cite{scherrer:statistics,sheth:non-linear}.  Both theoretical
models as well as observations suggest that dark matter caustics lead
to substructure inside halos {}\cite{kinney:evidence}.  Also recent
high--resolution $N$--body simulations suggest that 15\%--40\% of the
simulated halos show a significant amount of substructure (see
{}\cite{jing:density} and references therein).  To generalize the halo
model, the correlations inside the halo are taken into account.

Consider the expansion of $G_c[h|{\mathbf{y}}_i]-1$ in $h$:
\begin{align}
\lefteqn{G_c[h|{\mathbf{y}}_i]-1 = \int_{{\mathbb{R}}^d}{\rm d}{\mathbf{x}}_1\ h({\mathbf{x}}_1)\   
c_{[1]}({\mathbf{x}}_1|{\mathbf{y}}_i)+} \nonumber\\
& +\frac{1}{2!}\int_{{\mathbb{R}}^d}{\rm d}{\mathbf{x}}_1\int_{{\mathbb{R}}^d}{\rm d}{\mathbf{x}}_2\ h({\mathbf{x}}_1)h({\mathbf{x}}_2)
\times\nonumber\\
& \qquad \times\Big(c_{[1]}({\mathbf{x}}_1|{\mathbf{y}}_i)c_{[1]}({\mathbf{x}}_2|{\mathbf{y}}_i)  +
c_{[2]}({\mathbf{x}}_1,{\mathbf{x}}_2|{\mathbf{y}}_i) \Big) + \nonumber\\
& +\frac{1}{3!}\int_{{\mathbb{R}}^d}{\rm d}{\mathbf{x}}_1\int_{{\mathbb{R}}^d}{\rm d}{\mathbf{x}}_2\int_{{\mathbb{R}}^d}{\rm d}{\mathbf{x}}_3
\ h({\mathbf{x}}_1)h({\mathbf{x}}_2)h({\mathbf{x}}_3)\times\nonumber\\
& \qquad \times\Big(c_{[1]}({\mathbf{x}}_1|{\mathbf{y}}_i)c_{[1]}({\mathbf{x}}_2|{\mathbf{y}}_i)c_{[1]}({\mathbf{x}}_3|{\mathbf{y}}_i) + \nonumber\\ 
& \qquad +3\ c_{[2]}({\mathbf{x}}_1,{\mathbf{x}}_2|{\mathbf{y}}_i)c_{[1]}({\mathbf{x}}_3|{\mathbf{y}}_i) +
c_3({\mathbf{x}}_1,{\mathbf{x}}_2,{\mathbf{x}}_3|{\mathbf{y}}_i) \Big) + \nonumber\\
& + O[h^4] .
\end{align}
After inserting this expansion into Eq.~\eqref{eq:halo-expansion} and
collecting terms proportional to powers of $\varrho{}h(\cdot)$, with
the mean number density $\varrho=\varrho_p\mu$, one can directly
compare with the expansion~\eqref{eq:cumulant-exp} and read off the
correlation functions:
\begin{multline}
\xi_2({\mathbf{x}}_1,{\mathbf{x}}_2) = \\
=\frac{1}{\varrho_p\mu^2} \int_{{\mathbb{R}}^d}{\rm d}{\mathbf{y}}_1 
\Big(c_{[1]}({\mathbf{x}}_1|{\mathbf{y}}_1)c_{[1]}({\mathbf{x}}_2|{\mathbf{y}}_1)+c_{[2]}({\mathbf{x}}_1,{\mathbf{x}}_2|{\mathbf{y}}_1)\Big)+ \\
+ \frac{1}{\mu^2} \int_{{\mathbb{R}}^d}{\rm d}{\mathbf{y}}_1 \int_{{\mathbb{R}}^d}{\rm d}{\mathbf{y}}_2\
\xi_2^{(p)}({\mathbf{y}}_1,{\mathbf{y}}_2)\ c_{[1]}({\mathbf{x}}_1|{\mathbf{y}}_1)c_{[1]}({\mathbf{x}}_2|{\mathbf{y}}_2) ,
\end{multline}
\begin{align}
\label{eq:xi3_halomodel}
\lefteqn{\xi_3({\mathbf{x}}_1,{\mathbf{x}}_2,{\mathbf{x}}_3) = }\nonumber\\
=&\frac{1}{\varrho_p^2\mu^3}\int_{{\mathbb{R}}^d}{\rm d}{\mathbf{y}}_1 
\Big(c_{[1]}({\mathbf{x}}_1|{\mathbf{y}}_1)c_{[1]}({\mathbf{x}}_2|{\mathbf{y}}_1)c_{[1]}({\mathbf{x}}_3|{\mathbf{y}}_1)+\nonumber\\
&\qquad +3c_{[1]}({\mathbf{x}}_1|{\mathbf{y}}_1)c_{[2]}({\mathbf{x}}_2,{\mathbf{x}}_3|{\mathbf{y}}_1)+
c_3({\mathbf{x}}_1,{\mathbf{x}}_2,{\mathbf{x}}_3|{\mathbf{y}}_1)\Big) +\nonumber\\
& + \frac{3}{\varrho_p\mu^3} \int_{{\mathbb{R}}^d}{\rm d}{\mathbf{y}}_1 \int_{{\mathbb{R}}^d}{\rm d}{\mathbf{y}}_2\
\xi_2^{(p)}({\mathbf{y}}_1,{\mathbf{y}}_2)\times \nonumber\\
&\qquad\times\Big(c_{[1]}({\mathbf{x}}_1|{\mathbf{y}}_1)c_{[1]}({\mathbf{x}}_2|{\mathbf{y}}_2)c_{[1]}({\mathbf{x}}_3|{\mathbf{y}}_2)+
\nonumber\\
& \qquad\qquad +  
c_{[1]}({\mathbf{x}}_1|{\mathbf{y}}_1)c_{[2]}({\mathbf{x}}_2,{\mathbf{x}}_3|{\mathbf{y}}_2)\Big) +
\nonumber\\
& + 
\frac{1}{\mu^3}\int_{{\mathbb{R}}^d}{\rm d}{\mathbf{y}}_1\int_{{\mathbb{R}}^d}{\rm d}{\mathbf{y}}_2\int_{{\mathbb{R}}^d}{\rm d}{\mathbf{y}}_3\
\xi_3^{(p)}({\mathbf{y}}_1,{\mathbf{y}}_2,{\mathbf{y}}_3)\times \nonumber\\  
& \qquad\qquad \times c_{[1]}({\mathbf{x}}_1|{\mathbf{y}}_1)c_{[1]}({\mathbf{x}}_2|{\mathbf{y}}_2)c_{[1]}({\mathbf{x}}_3|{\mathbf{y}}_3).
\end{align}
Similarly the higher $n$--point correlation functions can be calculated.

In current calculations of the two-- and three--point functions for
the halo model {}\cite{sheth:non-linear,ma:halo} the galaxies inside
the halos are modeled as an inhomogeneous (finite) Poisson process.
The halo profile $c_{[1]}({\mathbf{x}}|{\mathbf{y}})$ is conditional on the cluster
center ${\mathbf{y}}$, but no substructure inside halos is present, i.e.\
$c_{[n]}=0$ for $n\ge2$.  In this case the above expressions simplify
to the result of {}\cite{scherrer:statistics}.

The simulation of such a point distribution can be carried out in a
multi--step approach similar to the simulation of the Gauss--Poisson
process (Appendix~\ref{sect:simulate-Gauss}).  First generate the
correlated cluster centers, e.g.\ by using a Gauss--Poisson process or
a low--resolution simulation, and then attach a secondary point process
either modeled as an inhomogeneous Poisson or $n$--point Poisson
cluster process.

\subsection{Halo substructure}

The following discussion shall serve mainly as an illustration of how
to incorporate halo substructure in calculations of the correlation
function.  To keep things simple the following assumptions are made:
the halo profile
$c_{[1]}({\mathbf{x}}|{\mathbf{y}})=c_{[1]}(|{\mathbf{x}}-{\mathbf{y}}|)$
is independent from the mass of the halo, and
$c_{[2]}({\mathbf{x}}_1,{\mathbf{x}}_2|{\mathbf{y}})$ factors into
$c_{[1]}({\mathbf{x}}_1|{\mathbf{y}})c_{[1]}({\mathbf{x}}_2|{\mathbf{y}})\gamma(|{\mathbf{x}}_1-{\mathbf{x}}_2|)$,
as expected for locally isotropic substructures.
Let
$P^{(p)}(k)=\int\frac{{\rm d}{\mathbf{x}}}{(2\pi)^3}\xi_2^{(p)}(|{\mathbf{x}}|){\rm e}^{-i{\mathbf{k}}\cdot{\mathbf{x}}}$
be the power spectrum of the spatial distribution of the halo centers,
and let $\widetilde{c}_{[1]}(k)$ and $\widetilde{\gamma}(k)$ be the
Fourier transform of $c_{[1]}$ and $\gamma$ respectively.  The power
spectrum of the galaxy distribution in the generalized halo model is
then
\begin{multline}
P(k) = \frac{(2\pi)^3}{\varrho_p\mu^2} |\widetilde{c}_1(k)|^2 
+ \frac{(2\pi)^6}{\mu^2} |\widetilde{c}_{[1]}(k)|^2 P^{(p)}(k) + \\
+\frac{(2\pi)^3}{\varrho_p\mu^2} \int{\rm d}{\mathbf{k}}'\
|\widetilde{c}_{[1]}(k')|^2 \widetilde{\gamma}(|{\mathbf{k}}-{\mathbf{k}}'|).
\end{multline}
This first two terms are the result of {}\cite{scherrer:statistics},
the additional term accounts for halo--substructure and involves a
folding of $\widetilde{c}_{[1]}$ with $\widetilde{\gamma}$ in
Fourier--space.  Similar expressions can be derived from
Eq.~\eqref{eq:xi3_halomodel} for the bispectrum.
Quantitative predictions for the galaxy distribution, similar to the
investigations by {}\cite{sheth:non-linear}, will be the topic of
future work.

\section{Some open problems}
\label{sect:problems}

Our investigations rested on the assumption that the correlation
functions exist and that the expansions of the p.g.fl.\ converge.  In
this case the p.g.fl., and consequently the point process, is
determined completely by the correlation functions.  The first
assumption, the existence of the correlation functions (the factorial
cumulants), does not impose dramatic restrictions for the models.  In
classical systems the mean number of points ${\mathbb{E}}[N]$ as well as the
factorial moments ${\mathbb{E}}[N(N-1)\cdots(N-n+1)]$ should be finite in any
bounded domain.  For the $n$--point Poisson cluster processes,
discussed in the preceding sections, at maximum $n$ points reside in a
cluster, which are themselves distributed according to a Poisson
process with constant number density.  Clearly in such a simple
situation both assumptions are satisfied.  However, even for
physically well motivated models, the convergence of the expansion of
the p.g.fl.\ may not be guaranteed, although the point process itself
and the correlation functions are well--defined.

\begin{figure}
\begin{center}
\epsfig{file=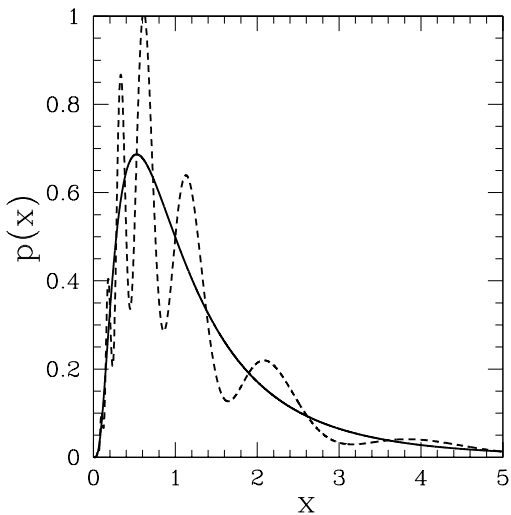,width=7cm}
\end{center}
\caption{\label{fig:lognormal}   The   probability   density  of   the
log--normal distribution (solid line, $\bar x=0$ and $\sigma=0.8$) and
a probability density  from the family of distributions  with the same
moments (dashed line, $\epsilon=0.5$ and $k=1$).}
\end{figure}

Perhaps the best known example of a probability distribution which is
not fully specified by its moments is the log--normal distribution.
The probability density of a log--normal random variable is given by
\begin{equation}
p(x) = \frac{1}{\sigma x \sqrt{2\pi}} 
\exp\Bigg(-\frac{(\log x-\bar x)^2}{2 \sigma^2}\Bigg)
\end{equation}
with parameters $\bar x$ and $\sigma^2$, the mean and variance of
$\log x$.  The moments (see Eq.~\eqref{eq:moment-randvar})
$m_n=\exp(n\bar x+\tfrac{1}{2}n^2\sigma^2)$ are well--defined, however
the expansion~\eqref{eq:phi-moment-expansion} of the characteristic
function is not convergent.  And indeed {}\cite{heyde:property} showed
that the probability density
\begin{equation}
p'(x) = p(x) \Bigg( 1+\epsilon 
\sin\Big(\frac{2\pi k}{\sigma^2} (\log x-\bar x)\Big) \Bigg)
\end{equation}
where  $0<\epsilon<1$  and $k$  is  a  positive  integer, has  moments
identical to the moments  of the log--normal distribution.  A comparison
of $p(x)$ and $p'(x)$ is shown in Fig.~\ref{fig:lognormal}.

A log--normal random field (an ``exponentiated'' Gaussian random
field) is positive at any point in space, and a point process can be
constructed using the value of the field as the local number density.
The multivariate log--normal distribution, and the log--normal random
field inherit the behaviour of the moments of the simple log--normal
distribution.  The point distribution obtained from the ``Poisson
sampled'' log--normal random field is not characterized completely by
its correlation functions as already discussed by
{}\cite{coles:lognormal}.  See also {}\cite{moller:log} for a
similar approach towards this ``log--Gaussian Cox process''.

In a Poisson cluster process (and also in the halo--model) the point
distribution inside the cluster is specified independently from the
distribution of the centers.  This constructive approach, and the
truncation of the moment expansion, guarantee the existence of these
processes.  A characterization result for the generalized Hermite
distribution, closely related to the general $n$--point process
considered in Sect.~\ref{sect:general-n-point}, is discussed by
{}\cite{milne:generalized}.  Also well--defined point processes which
do not impose such a truncation of the moment expansion are possible.
A simple model is the line--segment process used in
Sect.~\ref{sect:twod-example}, where the number of points per cluster
is a Poisson random variable.  Attempts towards a general
characterization of point processes were conducted by
{}\cite{ammann:structure,ammann:count,waymire:analysis}, and partially
succeeded for the case of infinitely divisible point processes.

One can show that any regular infinitely divisible point processes is
a Poisson cluster process (e.g.\ {}\cite{daley:introduction},
regular means that a cluster with an infinite number of points has
probability zero).  An infinitely divisible point processes may be
constructed as a superposition of any number of {\em independent}
point processes.  It is interesting to note that the log--normal
distribution is infinitely divisible {}\cite{thorin:infinite},
although the expansion of the characteristic function in terms of
moments {}\eqref{eq:phi-moment-expansion} does not converge.

On small scales the galaxy correlation function is scale invariant:
$\xi_2\propto r^{-\gamma}$.  If a cut--off at some large scales is
present, and the constraints for the density and the correlation
functions are satisfied, a model based on a Poisson cluster process
becomes feasible. Unfortunately, the superposition of {\em
independent} point processes, as implied by the infinite divisibility
of a Poisson cluster process, does not seem to be a good model
assumption for the interconnected network of {\em correlated} walls
and filaments, as observed in the galaxy distribution
(e.g.~\cite{huchra:cfa2s1}).
The correlation functions for the galaxy distribution are close to
zero for large separations, but from current observations one can not
infer a definite cut--off.  As discussed in Sect.~\ref{sect:physical}
for the Gauss--Poisson process, the large--scale behaviour of the
correlation functions plays an important role in the construction of
the Poisson cluster processes. Moreover, the dynamical equations
governing the evolution of large--scale structures are non--local (see
{}\cite{kofman:dynamics} and references therein).  Therefore it
seems worthwhile to consider also point process models which are not
infinitely divisible.
Unfortunately, beyond infinitely divisible point processes it is not
clear what kind of properties the correlation functions have and
especially what kind of additional constraints arise.

\section{Summary and conclusion}
\label{sect:summary}

The Gaussian random field, fully specified by the mean and its
correlation function, is one of the reference models employed in
cosmology.  Typical inflationary scenarios suggest that the primordial
mass--density field is a realization of a Gaussian random field.
Non--Gaussian features in the present day distribution of mass may be
attributed either to the non--linear process of structure formation,
or to a non--Gaussian primordial density field.  Observations of the
large--scale distribution of galaxies however provide us with a
distribution of points in space. The process of galaxy formation may
introduce further non--Gaussian features in the galaxy point
distribution. In this paper a direct approach towards the
characterization of this point set was pursued.  The statistical
properties of the point distribution can be specified by the sequence
of correlation functions $\xi_n$.  In close analogy to the Gaussian
random field, a Gaussian point distribution, the Gauss--Poisson point
process, was constructed.  This random point set is fully specified by
its mean number density $\varrho$, the two--point correlation function
$\xi_2(r)$, and $\xi_n=0$ for $n>2$.  Important constraints on
$\varrho$ and $\xi_2(r)$, not present for the Gaussian random field,
show up.  Namely, $\xi_2(r)\ge0$ for all $r$, and the variance of the
number of points must not exceed twice the value of a Poisson process.
The violation of these constraints indicates non--Gaussian features in
the galaxy distribution.
The equivalence of the Gauss--Poisson point process with a Poisson
cluster point process leads to a simple simulation algorithm for such
a point distribution.  Using the $J$--function, higher--order
correlations were detected in both a two--dimensional example and the
galaxy distribution.  The comparison with the Gauss--Poisson point
process allows us to quantify the level of significance of these
non--Gaussian features. Using these methods {}\cite{kerscher:reflex}
could show that the distribution of galaxy clusters may not be modeled
by a Gauss--Poisson process at a significance level of 95\%.

The formal approach based on the probability generating functional 
(p.g.fl.)  facilitated the definition, the characterization, and the 
simulation of the Gauss--Poisson point process.  The inclusion of 
higher--order correlation functions was straightforward, leading to the 
$n$--point Poisson cluster processes.  Both the definition and the 
simulation algorithm were detailed for the three--point Poisson 
cluster process.
The Gauss--Poisson point process and the two--point Poisson cluster 
process are equivalent.  However, this is not true for the $n$--point 
case $n>2$ anymore.  The set of general $n$--point processes, 
resulting from a truncation of the cumulant expansion of the p.g.fl.\ 
after the $n$--th order, contains all $n$--point Poisson cluster 
processes as a true subset.  This was discussed for the three--point 
case explicitly.
Although models based on the $n$--point Poisson cluster process are
not the most general ones, they cover a broad range of clustering
point distributions.  A Poisson cluster process can be simulated
easily and is especially helpful for comparing statistical methods
and estimators.

The inclusion of more and more points in the randomly placed clusters
is only one way to extend the Gauss--Poisson point process into the
strongly--correlated regime.  In the halo model one allows for
correlations between the cluster centers.  Typically the halo (i.e.\
the galaxy cluster) is modeled without substructure.  Again using the
p.g.fl., the influence of correlations inside a halo on the $n$--point
correlation functions of the resulting point distribution could be
calculated.

All the models discussed above offer some insight into certain aspects
of the clustering of the galaxy distribution.  As argued in the
preceding section, point process models which are not decomposable
into independent point processes seem more appropriate.
Unfortunately, even basic mathematical questions concerning the
(complete) characterization of these models in terms of moments and
beyond are still open.

\subsection*{Acknowledgments}

I would like to thank Lothar Heinrich, Klaus Mecke, Katja Schladitz,
Peter Sch{\"u}cker, Alex Szalay, and Herbert Wagner for suggestions
and helpful comments.  Especially I would like to thank Istvan Szapudi
for several interesting and beneficial discussions and Claus Beisbart
and Alvaro Dominguez for discussions and extensive comments on the
manuscript.  I acknowledge support from the NSF grant AST~9802980 and
from the {\em Sonderforschungsbereich 375 f{\"u}r Astroteilchenphysik
der DFG}.

\appendix

\section{Characteristic and generating functions of random variables}
\label{sect:random-variables}

A  short  review  dealing  with  the  characteristic  and  probability
generating {\em  function} (p.g.f.)  of  a {\em random  variable} and
their expansions  in terms  of moments, cumulants,  factorial moments,
and factorial cumulants is given.   This Appendix is meant to serve as
an illustration  highlighting the analogies between  expansions of the
probability generating {\em  functional} (p.g.fl.) and the probability
generating {\em function} (see also {}\cite{daley:introduction}).
To keep  this summary simple it is assumed  that the moments
etc.\  exist  and  the  expansions  converge.   For  a  more  thorough
treatment  of  characteristic   and  generating  functions  see  e.g.\
{}\cite{lukacs:survey,kendall:advanced1,linnik:decomposition}.

The moments of a random variable $x$ with probability distribution $F$
are defined by
\begin{equation}
\label{eq:moment-randvar}
m_{k} = {\mathbb{E}}\big[x^k\big]  = \int {\rm d} F(x)\ x^k .
\end{equation}
If $n$ is a discrete random variable, especially if $n$ is
integer--valued and greater equal zero, it is often more convenient to
work with the factorial moments:
\begin{equation}
m_{[k]} = {\mathbb{E}}\big[n^{[k]}\big]  = \int {\rm d} F(n)\ n^{[k]} 
= \sum_{n=0}^\infty p_n\ n^{[k]},
\end{equation}
where $n^{[k]}=n(n-1)\cdots(n-k+1)$, and $p_n$ is the probability that
the random variable takes the value $n$. Similarly, for point
processes it is more convenient to work with product densities (or
factorial moment measures), instead of moment measures.

The characteristic function $\varphi(t)$, $t\in{\mathbb{R}}$ of a distribution $F$
is defined as
\begin{equation}
\label{eq:def-phi}
\varphi(s) = \int_{-\infty}^\infty {\rm d} F(x)\ {\rm e}^{isx} ,
\end{equation}
and serves  as a  generating function for  the moments.  Expanding the
exponential one can easily verify that
\begin{equation}
m_k = (-i)^k \frac{{\rm d}^k\varphi(s)}{{\rm d} s^k} \Big|_{s=0} .
\end{equation}
By inverting  one obtains  the expansion of  $\varphi(s)$ in  terms of
moments:
\begin{equation}
\label{eq:phi-moment-expansion}
\varphi(s) =  \sum_{k=0}^\infty \frac{(is)^k}{k!} m_k .
\end{equation}

The  probability  generating  function  (p.g.f.) $P(z)$  of  a  random
variable $n$ is defined as
\begin{equation}
P(z) = {\mathbb{E}}[z^n] .
\end{equation}
Note that $P({\rm e}^{it})=\varphi(t)$.
For a nonnegative integer--valued random variable one obtains the
expansions
\begin{align}
P(z) &= \sum_{k=0}^\infty p_k\ z^k ,\\
P(1+t) &= 1+\sum_{k=1}^\infty  \frac{t^k}{k!} m_{[k]},
\end{align}
in terms of the probabilities $p_n$.  $P(1+t)$ serves as the 
generating function for the factorial moments $m_{[k]}$.
Similarly, the product densities (factorial moment measures) for a point 
process can be derived as functional derivatives (Frechet 
derivatives) of the probability generating functional.

Using  $P({\rm e}^{it})=\varphi(t)$  one  can  derive the  relation  between
moments and factorial moments:
\begin{align}
\label{eq:moment-fact-moment}
\varphi(s) & = \sum_{k=0}^\infty \frac{(is)^k}{k!} m_k = P({\rm e}^{is}) 
\nonumber \\
& = 1+\sum_{t=0}^\infty \frac{(is)^t}{t!}\sum_{k=1}^\infty m_{[k]}\ s_2(t,k),
\end{align}
where
\begin{equation}
\label{eq:def-sterling}
s_2(t,k) = \sum_{l=1}^k (-1)^{k-l}\frac{l^t}{(k-l)!\ l!}
\end{equation}
are the Stirling numbers of the second kind, the number of partitions
which split the set $\{1,\ldots,t\}$ into $k$ pairwise disjoint
nonempty sets (see e.g.\ {}\cite{wilf:generatingfunctionology}).
Since  $s_2(t,k)=0$  for  $k>t$,   which  is  also  respected  by  the
expression~\eqref{eq:def-sterling}, one finally arrives at
\begin{equation}
m_t = \sum_{k=1}^t s_2(t,k)\  m_{[k]} .
\end{equation}

One considers  not only  the moments $m_k$  of a random  variable, but
also the cumulants $c_k$ defined by
\begin{equation}
\label{eq:cumulant-randvar}
\exp\Bigg(\sum_{k=1}^\infty \frac{(is)^k}{k!} c_k\Bigg) = 
1+\sum_{k=1}^\infty \frac{(is)^k}{k!} m_k = \varphi(s).
\end{equation}
Clearly,  $\log\varphi(s)$ serves  as a  generating function  for the
cumulants $c_k$.  Perhaps the best known cumulant is the variance
\begin{equation}
c_2 = \sigma^2 = m_2 - m_1^2.
\end{equation}
Similarly,  for   nonnegative  integer  valued   random  variable  the
factorial cumulants $c_{[k]}$ are defined by
\begin{equation}
\label{eq:fact-cumulant-randvar}
\exp\Bigg(\sum_{k=1}^\infty \frac{t^k}{k!} c_{[k]}\Bigg) = 
1+\sum_{k=1}^\infty \frac{t^k}{k!} m_{[k]} = P(1+t).
\end{equation}
Hence, $\log P(1+t)$ serves as the generating function of the factorial
cumulants $c_{[k]}$.
The correlation functions $\xi_n$ used in cosmology, are the densities
of the normalized factorial cumulant measures of a point processes,
corresponding to the factorial cumulants $c_{[k]}$ of a discrete
random variable.
Setting  $t={\rm e}^{is}-1$  in Eq.~\eqref{eq:fact-cumulant-randvar}  and
comparing  term  by   term  with  Eq.~\eqref{eq:cumulant-randvar}  one
obtains the  same relation between cumulants  and factorial cumulants,
as between moments an factorial moments:
\begin{equation}
\label{eq:rv-cumulant-fact-cumulant}
c_k = \sum_{l=1}^k s_2(k,l)\ c_{[l]} .
\end{equation}

\section{Simulation algorithms}

\subsection{The Gauss--Poisson point process}
\label{sect:simulate-Gauss}

As discussed in Sect.~\ref{sect:gp-as-pc} every Gauss--Poisson process
is a Poisson cluster process and therefore can be simulated easily.
For a given number density $\varrho$ and a two-point correlation
function $\xi_2(r)$ fulfilling the constraints~\eqref{eq:const1} and
{}\eqref{eq:const2}, realizations of the Gauss--Poisson process can be
generated straightforwardly.  With the normalization
$C_2=\int_{{\mathbb{R}}^d}{\rm d}{\mathbf{x}}\ \xi_2(|{\mathbf{x}}|)$ and $\int_{{\mathbb{R}}^d}{\rm d}{\mathbf{x}}\
f(|{\mathbf{x}}|)=1$ one calculates the quantities needed for the simulation:
$f(r)=\xi_2(r)/C_2$, $q_2=\frac{\varrho C_2}{2-\varrho C_2}$,
$q_1=1-q_2$, and $\varrho_p=\varrho(1-\varrho C_2/2)$.  The
constraint~\eqref{eq:const2-alternate} now can be written as
$C_2\varrho\le1$.  The simulation is carried out in two steps:
\begin{itemize}
\item  First  generate the  cluster  centers  according  to a  Poisson
distribution with number density $\varrho_p$.
\item For each  cluster center ${\mathbf{x}}$ draw a  uniform random number $z$
in $[0,1]$. If $z<q_1$, then keep only the point ${\mathbf{x}}$.  If $z\ge q_1$
then keep the point ${\mathbf{x}}$ and additionally chose a random direction on
the unit  sphere and a distance  $d$ with the  probability density $f$
and place the second point according to them.
\end{itemize}
To get the correct point pattern inside a given window, one also has
to use cluster centers outside the window to ensure that any possible
secondary point inside the window is included.

\subsection{The three--point Poisson cluster process}
\label{sect:simul-threepoint}

In the following the algorithm for the simulations of the three--point
cluster process is given.  The expressions~\eqref{eq:f-xi-three}
together with the normalization conditions for $f_2$ and $f_3$ serve
as a starting point.  An algorithm similar to the one for the
Gauss--Poisson process described in Appendix.~\ref{sect:simulate-Gauss}
can be constructed:

Given  the number density  $\varrho$, and  the two--  and three--point
correlation functions, $\xi_2$ and $\xi_3$, one defines
\begin{equation}
C_2 = \int_{{\mathbb{R}}^d}{\rm d}{\mathbf{x}}\ \xi_2(|{\mathbf{x}}|) 
\text{ and }
C_3 = \int_{{\mathbb{R}}^d}{\rm d}{\mathbf{x}}\int_{{\mathbb{R}}^d}{\rm d}{\mathbf{y}}\ \xi_3(0,{\mathbf{x}},{\mathbf{y}}) .
\end{equation}
Using the normalization of $f_2$ and $f_3$ one obtains
\begin{align}
q_2 &= \frac{3(C_2\varrho-C_3\varrho^2)}{(6-3C_2\varrho+C_3\varrho^2)} ,
\nonumber\\
q_3 &= \frac{C_3\varrho^2}{(6-3C_2\varrho+C_3\varrho^2)} ,\\
\varrho_p &= \frac{\varrho}{6}(6-3C_2\varrho+C_3\varrho^2), \nonumber\\
q_1 &= 1-q_2-q_3 ,\nonumber
\end{align}
resulting in 
\begin{align}
f_2({\mathbf{x}}_1,{\mathbf{x}}_2) & = \frac{1}{C_2-C_3\varrho} \Big(\xi_2({\mathbf{x}}_1,{\mathbf{x}}_2) -\nonumber\\
&\qquad\qquad -\varrho\int_{{\mathbb{R}}^d}{\rm d}{\mathbf{x}}_3\ \xi_3({\mathbf{x}}_1,{\mathbf{x}}_2,{\mathbf{x}}_3)\Big) , \nonumber\\
f_2^{(3)}({\mathbf{x}}_1,{\mathbf{x}}_2) & = \frac{1}{C_3}\ 
\int_{{\mathbb{R}}^d}{\rm d}{\mathbf{x}}_3\ \xi_3({\mathbf{x}}_1,{\mathbf{x}}_2,{\mathbf{x}}_3),\\
f_3({\mathbf{x}}_1,{\mathbf{x}}_2,{\mathbf{x}}_3) & = \frac{1}{C_3}\ \xi_3({\mathbf{x}}_1,{\mathbf{x}}_2,{\mathbf{x}}_3).\nonumber
\end{align}
Since  $\varrho_c$ and the  $q_n$ are  positive numbers,  the constraints
$C_3\varrho^2\le C_2\varrho\le 2+C_3\varrho^2/3$ must be satisfied and
the relation $\xi_2=C_2f_2+C_3\varrho(f_2-f_2^{(3)})$ holds.
The algorithm now reads:
\begin{itemize}
\item  First  generate the  cluster  centers  according  to a  Poisson
distribution with number density $\varrho_p$.
\item For each  cluster center ${\mathbf{x}}$ draw a  uniform random number $z$
in $[0,1]$.  If  $z<q_1$, then keep only the  point ${\mathbf{x}}$.  If $q_1\le
z<q_1+q_2$ then keep  the point ${\mathbf{x}}$ and additionally  chose a random
point according  to the probability density $f_2$.   If $q_1+q_2\le z$
then  keep the  point ${\mathbf{x}}$,  chose a  second point  according  to the
probability density $f_2^{(3)}$, and a third point according to $f_3$.
\end{itemize}

\section{Cumulants and factorial cumulants}
\label{sect:cumulants-factorial-cumulants}

Consider the expansion~{}\eqref{eq:cumulant-exp} of the p.g.fl.\ in
terms of the factorial cumulants $c_{[k]}({\mathbf{x}}_1,\ldots,{\mathbf{x}}_k)$:
\begin{equation}
\begin{split}
\log G[h] & =  \sum_{k=1}^\infty \frac{1}{k!}
\int_{{\mathbb{R}}^d}{\rm d}{\mathbf{x}}_1\cdots\int_{{\mathbb{R}}^d}{\rm d}{\mathbf{x}}_k \\ 
& \qquad c_{[k]}({\mathbf{x}}_1,\ldots,{\mathbf{x}}_k) (h_1-1)\cdots(h_k-1) \\
& = \sum_{k=1}^\infty 
\frac{1}{k!}\int_{{\mathbb{R}}^d}{\rm d}{\mathbf{x}}_1\cdots\int_{{\mathbb{R}}^d}{\rm d}{\mathbf{x}}_k\ c_{[k]}({\mathbf{x}}_1,\ldots,{\mathbf{x}}_k)\\ 
& \qquad
\bigg\{(-1)^k +\sum_{n=1}^k(-1)^{k-n} \sum_{I\in J_n^k} h_{I_1}\cdots h_{I_n}
\bigg\}, 
\end{split}
\end{equation}
with $h_1=h({\mathbf{x}}_1)$ and $J_n^k$ is formed by the ordered subsets of
$\{1,\ldots,k\}$ with $n\le k$ distinct entries. Hence, a subset $I\in
J_n^k$ consists out of $n$ distinct numbers $\{I_1,\ldots,I_n\}$ with
$I_1<\ldots<I_n$, e.g\ $J_2^3=\{\{1,2\},\{1,3\},\{2,3\}\}$.
Using $\Phi[h]=G[{\rm e}^{ih}]$ one obtain 
\begin{equation}
\begin{split}
\lefteqn{\log\Phi[h] = 
\sum_{k=1}^\infty \frac{1}{k!} 
\int_{{\mathbb{R}}^d}{\rm d}{\mathbf{x}}_1\cdots\int_{{\mathbb{R}}^d}{\rm d}{\mathbf{x}}_k\ 
c_{[k]}({\mathbf{x}}_1,\ldots,{\mathbf{x}}_k)\times}\\
& \times \bigg\{(-1)^k+\sum_{n=1}^k(-1)^{k-n}\sum_{I\in J_n^k}
\exp\big(i(h_{I_1}+\ldots+h_{I_n})\big) \bigg\}, \\
= & \sum_{k=1}^\infty \frac{1}{k!} 
\int_{{\mathbb{R}}^d}{\rm d}{\mathbf{x}}_1\cdots\int_{{\mathbb{R}}^d}{\rm d}{\mathbf{x}}_k\ 
c_{[k]}({\mathbf{x}}_1,\ldots,{\mathbf{x}}_k)\times\\
& \times \bigg\{(-1)^k+   
\sum_{m=0}^\infty \frac{i^m}{m!}\sum_{n=1}^k(-1)^{k-n}\sum_{I\in J_n^k}
(h_{I_1}+\ldots+h_{I_n})^m \bigg\}.
\end{split}
\end{equation}
In the $m$ sum  the  first term  with  $m=0$ equals  $-(-1)^k$,
canceling with the $(-1)^k$ inside the braces:
\begin{equation}
\label{eq:cumulant-factorial-final}
\begin{split}
\lefteqn{\log\Phi[h] = 
\sum_{k=1}^\infty \frac{1}{k!} 
\int_{{\mathbb{R}}^d}{\rm d}{\mathbf{x}}_1\cdots\int_{{\mathbb{R}}^d}{\rm d}{\mathbf{x}}_k\ 
c_{[k]}({\mathbf{x}}_1,\ldots,{\mathbf{x}}_k)\times}\\
& \times \bigg\{
\sum_{m=1}^\infty \frac{i^m}{m!}\sum_{n=1}^k(-1)^{k-n}\sum_{I\in J_n^k}
(h_{I_1}+\ldots+h_{I_n})^m \bigg\}\\
= & \sum_{m=1}^\infty \frac{i^m}{m!} \Bigg[ 
\sum_{k=1}^\infty \frac{1}{k!} 
\int_{{\mathbb{R}}^d}{\rm d}{\mathbf{x}}_1\cdots\int_{{\mathbb{R}}^d}{\rm d}{\mathbf{x}}_k\ 
c_{[k]}({\mathbf{x}}_1,\ldots,{\mathbf{x}}_k)\times\\
& \times \bigg\{
\sum_{n=1}^k(-1)^{k-n}\sum_{I\in J_n^k}
(h_{I_1}+\ldots+h_{I_n})^m \bigg\}\Bigg] .\\
\end{split}
\end{equation}
The expression inside the braces $[\cdots]$
in~\eqref{eq:cumulant-factorial-final} equals
\begin{multline}
\int_{{\mathbb{R}}^d}{\rm d}{\mathbf{x}}_1\cdots\int_{{\mathbb{R}}^d}{\rm d}{\mathbf{x}}_m\
c_k({\mathbf{x}}_1,\ldots,{\mathbf{x}}_m)\ h_1\cdots h_m.
\end{multline}
This fixes the relation between the cumulants and factorial
cumulants. However, there is no straightforward way to simplify this
expression.  Above all the theorem of Marcinkiewicz demands that as
soon as $k>2$, the $k$--sum always has to be an infinite sum (see
Sect.~\ref{sect:marcinkiewicz}). Hence, the cumulants $c_k(\cdot)$
depend on an infinite alternating sum of the factorial cumulants
$c_{[k]}(\cdot)$, and vice versa.


\end{document}